\documentclass[aps,showpacs,preprintnumbers,amsmath, amssymb,nofootinbib]{revtex4}

\usepackage{float}
\usepackage{graphics,epsfig}
\usepackage{bm}
\usepackage{slashed}
\usepackage{graphicx}
\usepackage{amsmath}
\usepackage{amsfonts}
\usepackage{amssymb}
\usepackage{color}%
\usepackage{dcolumn}

\begin{document}
\baselineskip=0.8 cm \title{Dynamical gap from
holography in the charged dilaton black hole}
\author{Xiao-Mei Kuang$^{1}$}
\author{Bin Wang$^{1}$}
\author{Jian-Pin Wu$^{2,3,4}$}
\affiliation{$^{1}$INPAC, Department of Physics and Shanghai
Key Lab for Particle Physics and Cosmology,
Shanghai Jiao Tong University, Shanghai 200240,
China\\
$^{2}$Department of Physics, Hanyang University, Seoul 133-791, Korea\\
$^{3}$Center for Quantum Spacetime, Sogang University, Seoul 121-742, Korea\\
$^{4}$Department of Physics, School of Mathematics and Physics, Bohai University, JinZhou, 121013, China }

\vspace*{0.2cm}
\begin{abstract}
\baselineskip=0.6 cm
\begin{center}
{\bf Abstract}
\end{center}
~~We study the holographic non-relativistic
fermions in the presence of bulk dipole coupling
in charged dilatonic black hole background. We
explore the nontrivial effects of the bulk
dipole coupling, the fermion charge as well as
 the dilaton field on the flat band, the Fermi
 surface and the emergence of the gap by investigating
 the spectral function of the non-relativistic fermion
 system. In particular, we find that the presence of the flat band
 in the non-relativistic case will suppress the Fermi momentum. 
 Besides, we observe that the effect of the dipole coupling in the dilaton gravity is more explicit. Finally,
 we consider the non-relativistic fermions at nonzero temperature.
 A phase transition from insulator to a conducting state is observed as the fermion system becomes hotter.
\end{abstract}

\maketitle
\newpage
\vspace*{0.2cm}

\section{Introduction}

The AdS/CFT correspondence is a powerful method
to investigate the strongly coupled many body
phenomena by relating certain interacting quantum
field theories to classical gravity systems. An
interesting application of such duality is the
study of condensed matter physics from the
remarkable connection in gravitational physics,
for reviews see for examples
\cite{Hartnoll,herzog,horowitz}.

Stimulated by the  AdS/CFT correspondence, the
simplest realization to construct gravitational
duals of the transition from normal state to
superconducting state is to deal with Einstein's
gravitational theory with a negative cosmological
constant coupled to a gauge field. This
idea caught some interesting properties of the
realistic superconductor by discussing holography in
the simplest charged AdS black hole background.
However, the RN-AdS black hole contains nonzero
entropy even at zero temperature.
This situation was improved by changing the
background bulk geometry to include a real scalar
field, the dilaton, which allows zero entropy at
zero temperature \cite{BHbackground1,BHbackground2,BHbackground3}.
At zero temperature, it was argued that states of matter can be
holographically described by a spacetime with an asymptotically
$AdS_4$ Einstein-Maxwell-dilaton theory, the quantum phase transitions
between different phases were exhibited \cite{Huijse}.
More holographic properties disclosed in the dilaton black holes can be found in  \cite{BHbackground2}.

In addition to modifying the bulk gravitational
backgrounds, recently there have been some
attempts to modify the boundary conditions in
describing the condensed matter system. In
\cite{D.Tong1}, it was shown that one can
implement the holographic non-relativistic
fermionic fixed points by imposing Lorentz
breaking boundary conditions instead of the
Lorentz covariant one on the Dirac spinor field,
which can lead to the presence of an infinite
flat band in the boundary field theory.  Further
studies on the holographic non-relativistic
fermionic fixed points have been reported in
\cite{D.Tong2,WJL1}. In \cite{WJL2,JPWu3}, the
holographic non-relativistic fermionic fixed
points were studied in the charged dilatonic
black hole and black brane.

To capture more general features and phenomena
relating to condensed matter physics, there have
been a lot of investigations to consider a
quantum field theory which contains fermions
charged under a global U(1) symmetry
\cite{Lee,HongLiuUniversality,HongLiuADS2,
HongLiuNonFermi,HongLiuSpinor,Cubrovic,wujianpin,Fang}.
However, many of these studies concentrated on
the fermions minimally coupled to gravity and
gauge fields. Recently, introducing the coupling
between the fermion and gauge field through a
dipole interaction in the bulk charged AdS black
hole background, it was remarkably found that as
the strength of the interaction is varied,
spectral density is transferred and beyond the
critical interaction strength a gap opens
up\cite{R.G.Leigh1}. The existence of Fermi
surfaces as the varying of the dipole coupling
was also disclosed\cite{R.G.Leigh2}. The
extensions of the investigation on the dipole
coupling to different bulk backgrounds were
reported in \cite{JPWu2,Wen,Kuang}.

In this work we will extend the study of the
dipole interaction to the charged dilatonic AdS
black hole background. Considering that different
from the charged AdS black hole, the charged
dilatonic AdS black hole has vanishing entropy in
the low temperature limit, it is a suitable bulk
background to study the holographic Fermi liquid.
Furthermore, the dilaton can couple directly not
only to the gauge field but also to the charged
scalar, which can help to further disclose the
role played by the dipole coupling in the
boundary theory of Fermi liquid. In our study,
instead of keeping Lorentz invariance for the
boundary theory, we will impose Lorentz violating
boundary terms for a spinor field following
\cite{D.Tong1}. We will investigate the
holographic spectral function behaviors at the
non-relativistic fermionic fixed points and
compare with the situations at the relativistic
fermonic fixed point\cite{Wen}. We will disclose
how the effects of dilaton, the dipole
coupling as well as the fermion charge modify the properties of Fermi gap,
Fermi momentum etc. in the Fermi system.

The organization of this paper is as follows. In
section II, we will derive the bulk Dirac
equations in 4-dimensional charged dilatonic AdS
black hole. Then we will briefly discuss the
holographic calculations of the retarded Green
functions of those fermionic operators for
non-relativistic theory. In section III, we will
present our numerical results for the
non-relativistic fixed point in different cases.
Finally, we will summarize our results in the last section.

\section{holographic setup}
In this section, we will derive the Dirac
equations of the bulk fermion coupling to the
gauge field through dipole interaction in the
charged dilatonic AdS black hole background. We
will employ the Lorentz violating boundary term
for the spinor field and study the holographic
fermionic systems through the perturbations on
the non-relativistic fermionic fixed point.

\subsection{Dirac equation}
We consider the non-minimal coupling between the
spin-1/2 fermions and the gauge field in the form
of the dipole interaction described by the bulk
action
\begin{eqnarray}
\label{actionspinor}
S_{bulk}=i\int d^{d}x \sqrt{-g}\overline{\zeta}\left(\Gamma^{a}\mathcal{D}_{a} - m - ip \slashed{F} \right)\zeta,
\end{eqnarray}
where $m$ and $p$ are the mass of the fermion
field and the dipole coupling parameter,
respectively. In the action,
$\Gamma^{a}=(e_{\mu})^{a}\Gamma^{\mu}$,
$\slashed{F}=\frac{1}{4}\Gamma^{\mu\nu}(e_\mu)^a(e_\nu)^bF_{ab}$
and
$\mathcal{D}_{a}=\partial_{a}+\frac{1}{4}(\omega_{\mu\nu})_{a}\Gamma^{\mu\nu}-iqA_{a}$,
with
$\Gamma^{\mu\nu}=\frac{1}{2}[\Gamma^\mu,\Gamma^\nu]$
and  the spin connection
$(\omega_{\mu\nu})_{a}=(e_\mu)^b\nabla_a(e_\nu)_b$,
where $(e_\mu)^{a}$ form a set of orthogonal
normal vector bases\cite{Conventions1}.

We intend to work in the charged dilatonic AdS
black hole background with the metric
\begin{eqnarray}\label{blackhole}
ds^2=-g_{tt}dt^2+g_{xx}(d\vec{x})^2+g_{rr}dr^2=e^{2B}[-fdt^2+(d\vec{x})^2]+\frac{1}{e^{2B}}\frac{dr^2}{f},~~~~F=dA,~~~~\phi=\frac{1}{2}\ln(1+\frac{Q}{r}), \nonumber \\
B=\ln\frac{r}{L}+\frac{3}{4}\ln(1+\frac{Q}{r}),~~~~ f=1-\frac{\nu L^2}{(Q+r)^3}, ~~~~ A_a=(\frac{\sqrt{3Q\nu}}{Q+r}-\frac{\sqrt{3Q}\nu^\frac{1}{6}}{L^\frac{2}{3}})(dt)_a,
\end{eqnarray}
which is a solution to the Einstein-Maxwell-Dilaton action in 4-dimensional spacetime
\cite{BHbackground1}
\begin{equation}
S_g=\frac{1}{16 \pi G}\int_Md^4x\sqrt{-g}[R-\frac{1}{4}e^\phi F_{ab}F^{ab}-\frac{3}{2}\nabla_a\phi\nabla^a\phi+\frac{6}{L^2}\cosh\phi].
\end{equation}
Here, $R$ and $L$ are the Ricci scalar and the
AdS radius, respectively. $\phi$ denotes the
dilaton field and $F=dA$ is the field strength of
$U(1)$ gauge field.
The temperature of the
charged dilaton black hole and chemical potential
near the boundary read
\begin{eqnarray}\label{tmu}
T=\frac{g_{tt}^{'}|_{r=r_+}}{4\pi}=\frac{3\nu^\frac{1}{6}}{4\pi L^\frac{5}{3}}\sqrt{r_+},~~~~~~
\mu=-\frac{\sqrt{3 Q} \nu^\frac{1}{6}}{L^\frac{2}{3}},
\end{eqnarray}
where $r_+$ is the black hole horizon with the
form $r_+=\nu^\frac{1}{3}L^\frac{2}{3}-Q$
obtained from $f({r_+})=0$. When
$\nu=\frac{Q^3}{L^2}$, the black hole has zero
temperature.

The Dirac equation
$(\mathcal{D}-m-ip\slashed{F})\zeta=0$ in the
bulk has the form
\begin{eqnarray}
\label{DiracEinFourier}
(\sqrt{g^{rr}}\Gamma^{r}\partial_{r}- m - \frac{i
p}{2} \sqrt{g^{rr}g^{tt}} \Gamma^{rt}
\partial_{r}A_{t})F -i(\omega+q
A_{t})\sqrt{g^{tt}}\Gamma^{t}F +i k
\sqrt{g^{xx}}\Gamma^{x}F=0,
\end{eqnarray}
after taking the ansatz $\zeta=(-g
g^{rr})^{-\frac{1}{4}}F e^{-i\omega t
+ik_{i}x^{i}}$ and setting $k_i=k\delta^i_{1}$
without loss of generality. It is convenient to
express $F$ into $F=(F_{1},F_{2})^{T}$ and choose
the following basis for our gamma
matrices\cite{Photoemission}
\begin{eqnarray}
\label{GammaMatrices}
\Gamma^r=\left( \begin{array}{cc}-\sigma^3 & 0 \\0 & -\sigma^3 \\\end{array}\right),
\ \ \Gamma^t=\left(\begin{array}{cc}i\sigma^1 & 0 \\0 & i\sigma^1 \\\end{array}\right),
\ \ \Gamma^1=\left(\begin{array}{cc}-\sigma^2 & 0 \\0 & \sigma^2 \\\end{array}\right),
\ \ \Gamma^2=\left(\begin{array}{cc}0 & -i\sigma^2 \\i\sigma^2 & 0 \\\end{array}\right).
\end{eqnarray}
The Dirac equation can be rewritten into
\begin{eqnarray} \label{DiracEF}
\sqrt{g^{rr}}\partial_{r}\left( \begin{matrix} F_{1} \cr  F_{2} \end{matrix}\right)
+m\sigma^3\otimes\left( \begin{matrix} F_{1} \cr  F_{2} \end{matrix}\right)
=\sqrt{g^{tt}}(\omega+qA_{t})i\sigma^2\otimes\left( \begin{matrix} F_{1} \cr  F_{2} \end{matrix}\right)
\mp  k \sqrt{g^{xx}}\sigma^1 \otimes \left( \begin{matrix} F_{1} \cr  F_{2} \end{matrix}\right) \nonumber \\
-p \sqrt{g^{tt}g^{rr}}\partial_{r}A_{t}\sigma^1\
\otimes \left( \begin{matrix} F_{1} \cr  F_{2} \end{matrix}\right).
\end{eqnarray}
Furthermore, to decouple the equation of motion,
we take the decomposition
$F_{\pm}=\frac{1}{2}(1\pm \Gamma^{r})F$ with
\begin{eqnarray} \label{gammarDecompose}
F_{+}=\left( \begin{matrix} 0\cr \mathcal{B}_{1} \cr  0\cr \mathcal{B}_{2} \end{matrix}\right),~~~~
F_{-}=\left( \begin{matrix} \mathcal{A}_{1} \cr  0\cr \mathcal{A}_{2}\cr 0 \end{matrix}\right),~~~~
and~~~~F_{\alpha} \equiv \left( \begin{matrix} \mathcal{A}_{\alpha} \cr  \mathcal{B}_{\alpha} \end{matrix}\right),\alpha=1,2.
\end{eqnarray}
Under such decomposition, the Dirac equation (\ref{DiracEF}) can be divided into
\begin{eqnarray} \label{DiracEAB1}
(\sqrt{g^{rr}}\partial_{r}\pm m)\left( \begin{matrix} \mathcal{A}_{1} \cr  \mathcal{B}_{1} \end{matrix}\right)
=\pm(\omega+qA_{t})\sqrt{g^{tt}}\left( \begin{matrix} \mathcal{B}_{1} \cr  \mathcal{A}_{1} \end{matrix}\right)
-(k \sqrt{g^{xx}}+p \sqrt{g^{tt}g^{rr}}\partial_{r}A_{t})\left( \begin{matrix} \mathcal{B}_{1} \cr  \mathcal{A}_{1} \end{matrix}\right)
~,
\end{eqnarray}
\begin{eqnarray} \label{DiracEAB2}
(\sqrt{g^{rr}}\partial_{r}\pm m)\left( \begin{matrix} \mathcal{A}_{2} \cr  \mathcal{B}_{2} \end{matrix}\right)
=\pm(\omega+qA_{t})\sqrt{g^{tt}}\left( \begin{matrix} \mathcal{B}_{2} \cr  \mathcal{A}_{2} \end{matrix}\right)
+(k \sqrt{g^{xx}}-p \sqrt{g^{tt}g^{rr}}\partial_{r}A_{t}) \left( \begin{matrix} \mathcal{B}_{2} \cr  \mathcal{A}_{2} \end{matrix}\right)
~.
\end{eqnarray}

It is straightforward to reduce the above two
equations into the flow equation of
$\xi_{I}\equiv
\frac{\mathcal{A}_{I}}{\mathcal{B}_{I}}(I=1,2)$
\begin{eqnarray} \label{DiracEF1}
(\sqrt{f}e^{B}\partial_{r}+2m)\xi_{I}=\left[ v_{-} + (-1)^{I} e^{-B} k  \right]
+ \left[ v_{+} - (-1)^{I} k e^{-B}  \right]\xi_{I}^{2}~
\end{eqnarray}
where $v_{\pm}=\frac{e^{-B}}{\sqrt{f}}\left[\omega+q A_t\right]\pm p A_t^{'}$.

Near the AdS boundary, from (\ref{DiracEF}) we
see that the reduced Dirac field behaves as
\begin{eqnarray} \label{BoundaryBehaviour}
F_{I} \buildrel{r \to \infty}\over {\approx} a_{I}r^{-m L}\left( \begin{matrix} 1 \cr  0 \end{matrix}\right)
+b_{I}r^{m L}\left( \begin{matrix} 0 \cr  1 \end{matrix}\right),
\qquad
I = 1,2~.
\end{eqnarray}
Then the value $G_{I}=\frac{a_I}{b_I}$ can be expressed in the form
\begin{eqnarray} \label{GreenFBoundary}
G_I=\lim_{r\rightarrow\infty}r^{2m}\xi_I
\end{eqnarray}
Thus, we can read off $G_{I}$ by solving the flow equation (\ref{DiracEF1}) with the boundary conditions at the horizon\cite{WJL2}
\begin{eqnarray}\label{GatTip}
\xi_I=
\left\{
\begin{array}{rl}
i~~~~~~~~~~~~~~~~~~~~~~~ \mbox{for $\omega \neq 0$};\\
(-1)^I\mathtt{sign}(k)~~~~~\mbox{for $\omega =0$}.
\end{array}\right.
\end{eqnarray}

It is worth indicating that from the flow
equation (\ref{DiracEF1}) and the above boundary
conditions, we can find $G_{I}$ with the following
symmetries:
\begin{eqnarray}
G_{1}(\omega,k)&=&G_{2}(\omega,-k);\nonumber\\
G_{1}(\omega,k;q,p)&=&-G_{2}^{*}(-\omega,k;-q,-p).
\end{eqnarray}
Especially, when $m=0$, $G_{1}$ and $G_{2}$ satisfy the relation
\begin{equation}\label{easy}
G_1(\omega,k)=-\frac{1}{G_2(\omega,k)}.
\end{equation}

\subsection{Non-relativistic fermionic fixed point}
Most available works on the holographic fermionic
systems focus on the perturbations on the
relativistic fixed point by keeping the Lorentz
invariance for the boundary theory. In
\cite{D.Tong1}, the authors first considered the
boundary term by dropping the Lorentz invariance
while still keeping the $U(1)$ global symmetry
$\psi\rightarrow e^{i\theta}\psi$, the rotational
invariance as well as the scale invariance. In
this spirit, the boundary term to the bulk action
(\ref{actionspinor}) reads
\begin{eqnarray} \label{boundary2}
S_{bdy}&=&\frac{1}{2}\int_{\partial\mathcal{M}}d^3x\sqrt{-gg^{rr}}\bar{\zeta}\Gamma^1\Gamma^2\zeta.
\end{eqnarray}
This boundary term keeps the variational
principle well-defined.

The variation of the on-shell action for the
Dirac spinor has the form
\begin{eqnarray} \label{variation2}
\delta S_D=\delta S_{bulk}+\delta S_{bdy}=-\int d^3x(\delta B_{+}^\dag A_{+}+ B_{-}^\dag \delta A_{-}+A_{+}^\dag \delta B_{+}+ \delta A_{-}^\dag  B_{-})
\end{eqnarray}
with $(A_{+},A_{-})=\frac{1}{\sqrt{2}}(\mathcal
{A}_1+\mathcal {A}_2,\mathcal {A}_1-\mathcal
{A}_2)$ and
$(B_{+},B_{-})=\frac{1}{\sqrt{2}}(\mathcal
{B}_1+\mathcal {B}_2,\mathcal {B}_2-\mathcal
{B}_1)$.

We can extract two groups of fermionic source and
the dual operator, which are $(B_{+},A_{+})$ and
$(A_{-},B_{-})$. The dimensions of the operators
are $\frac{3}{2}+m$ and $\frac{3}{2}-m$,
respectively \cite{D.Tong1}.

According to the AdS/CFT dictionary, the retarded
Green function for the non-relativistic fermionic
fixed point can be defined as
\begin{eqnarray} \label{smatrix2}
\left(
\begin{array}{c}
A_{+} \\B_{-} \\
\end{array}
\right)=G_{R}\left(
\begin{array}{c}
B_{+} \\A_{-} \\
\end{array}
\right).
\end{eqnarray}
Following the analysis in \cite{WJL2}, the matric $G_{R}=\left(
\begin{array}{cc}
\frac{2G_1G_2}{G_1+G_2} & \frac{G_1-G_2}{G_1+G_2} \\
 \frac{G_1-G_2}{G_1+G_2} & \frac{-2}{G_1+G_2} \\
\end{array}
\right)$ is off-diagonal and its eigenvalue
$\lambda_{\pm}$ can be expressed in terms of
$G_{I}$ as
\begin{eqnarray} \label{green2}
\lambda_{\pm}=\frac{G_{1}G_{2}-1\pm\sqrt{1+G_{1}^2+G_{2}^2+G_{1}^2G_{2}^2}}{G_{1}+G_{2}}.
\end{eqnarray}
Thus, the spectral function has the form
\begin{equation}
A(\omega,k)=Tr[Im G_R]=\frac{2 G_1 G_2-2}{G_1+G_2}.
\end{equation}

\section{numerical results}
We numerically integrate the flow equation
(\ref{DiracEF1}) and read off the asymptotic
values to compute the  retarded Green functions
with the Lorentz violating boundary term in the
charged dilatonic AdS black hole background. We
will calculate the fermion spectral function and
also the density of states. We will investigate
the effects of the dipole coupling, the fermion charge
and the dilaton field
on the holographic fermionic
system.

\subsection{Zero temperature}
In this subsection, we will show our numerical
results of the massless fermion in the limit of
zero temperature. For convenience, we fix $L=1$.
The chemical potential reads $\mu=-\sqrt{3}Q$.

\subsubsection{The dipole effect on the flat band and the Fermi surface }
\begin{figure}[ht]
\centering
\includegraphics[scale=0.3]{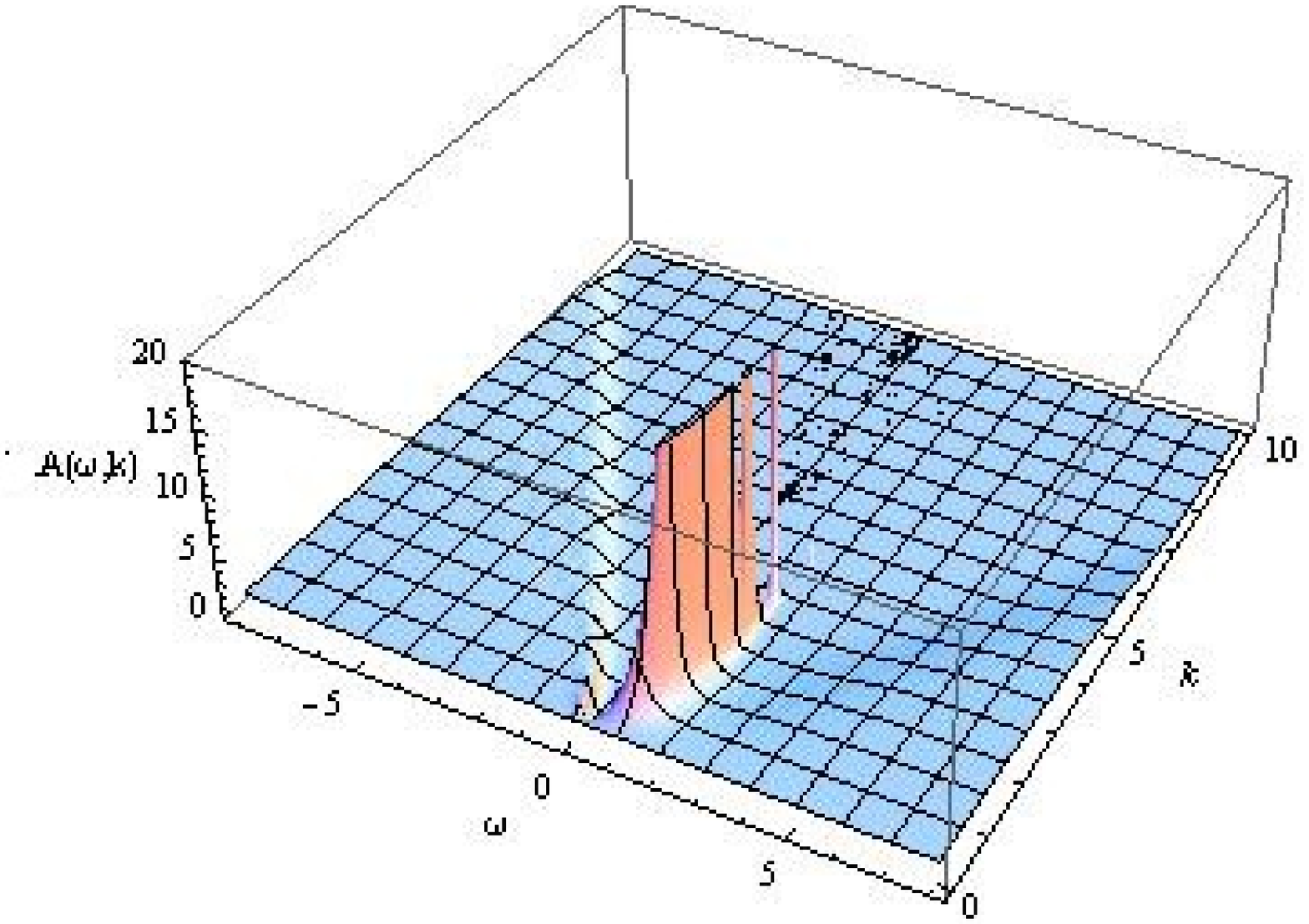}\hspace{1.5cm}
\includegraphics[scale=0.23]{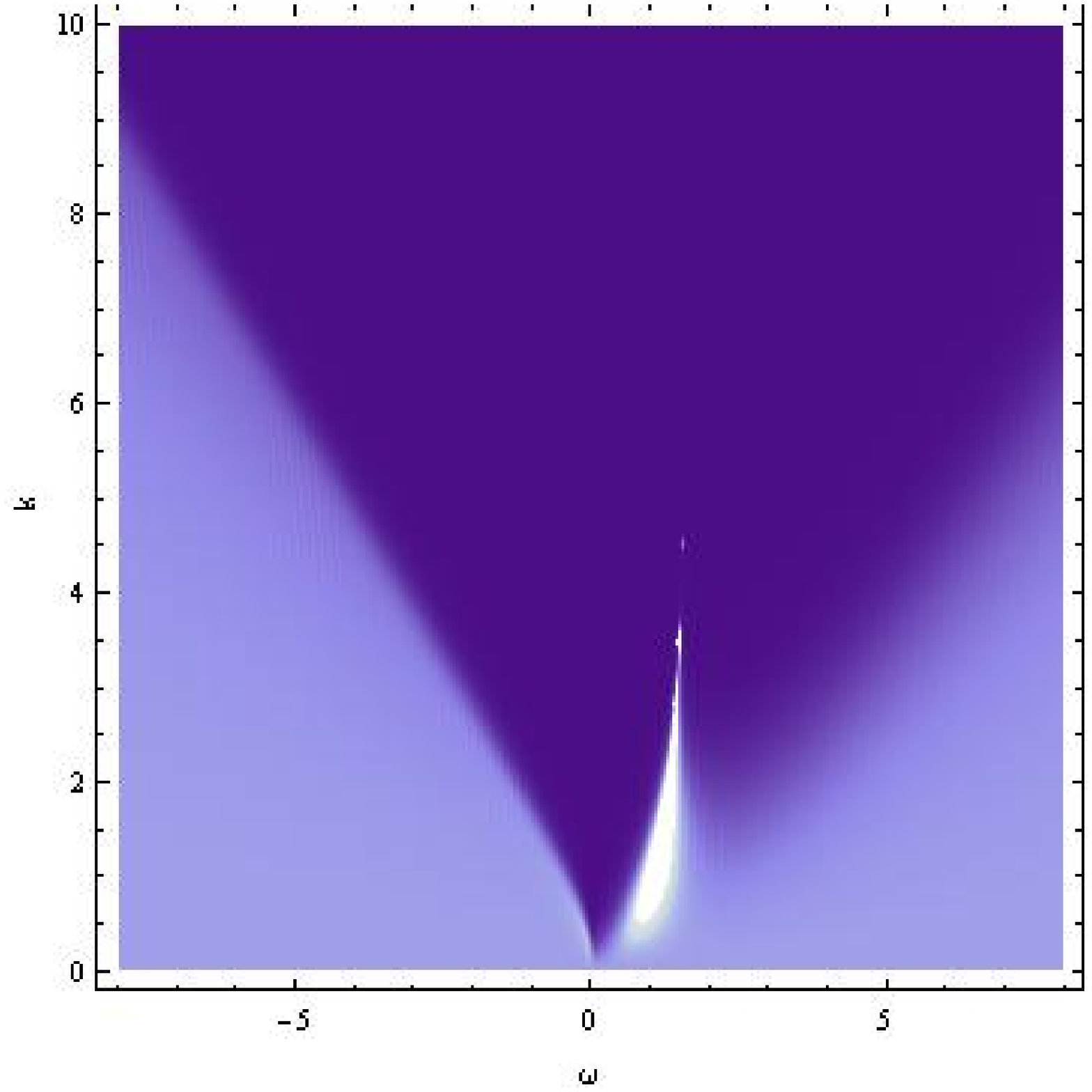}
\caption{The plots of $A(\omega,k)$ for the case
of $q=1$ and $p=0$.} \label{p0}
\end{figure}

\begin{figure}[ht]
\centering
\includegraphics[scale=0.3]{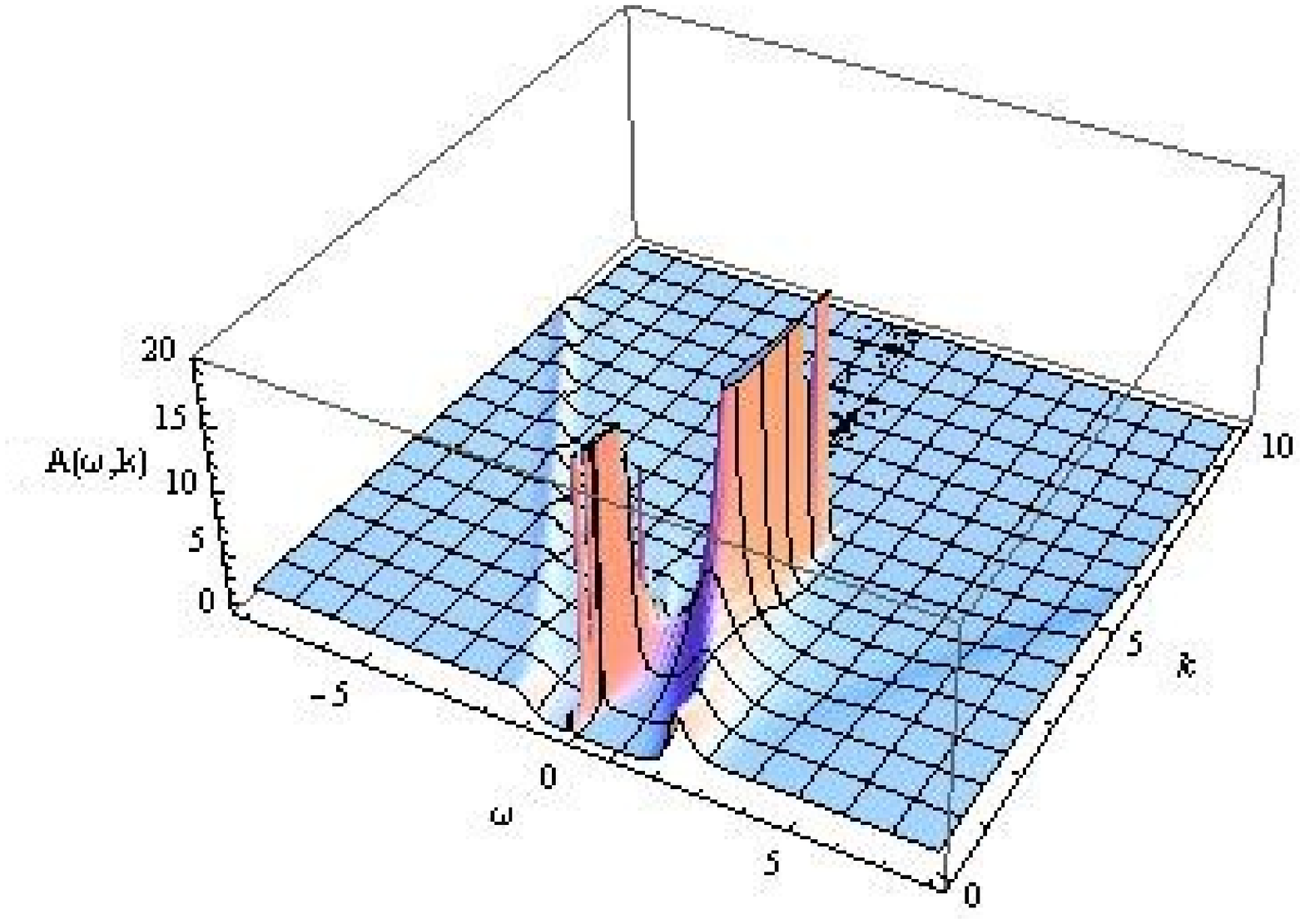}\hspace{1.5cm}
\includegraphics[scale=0.23]{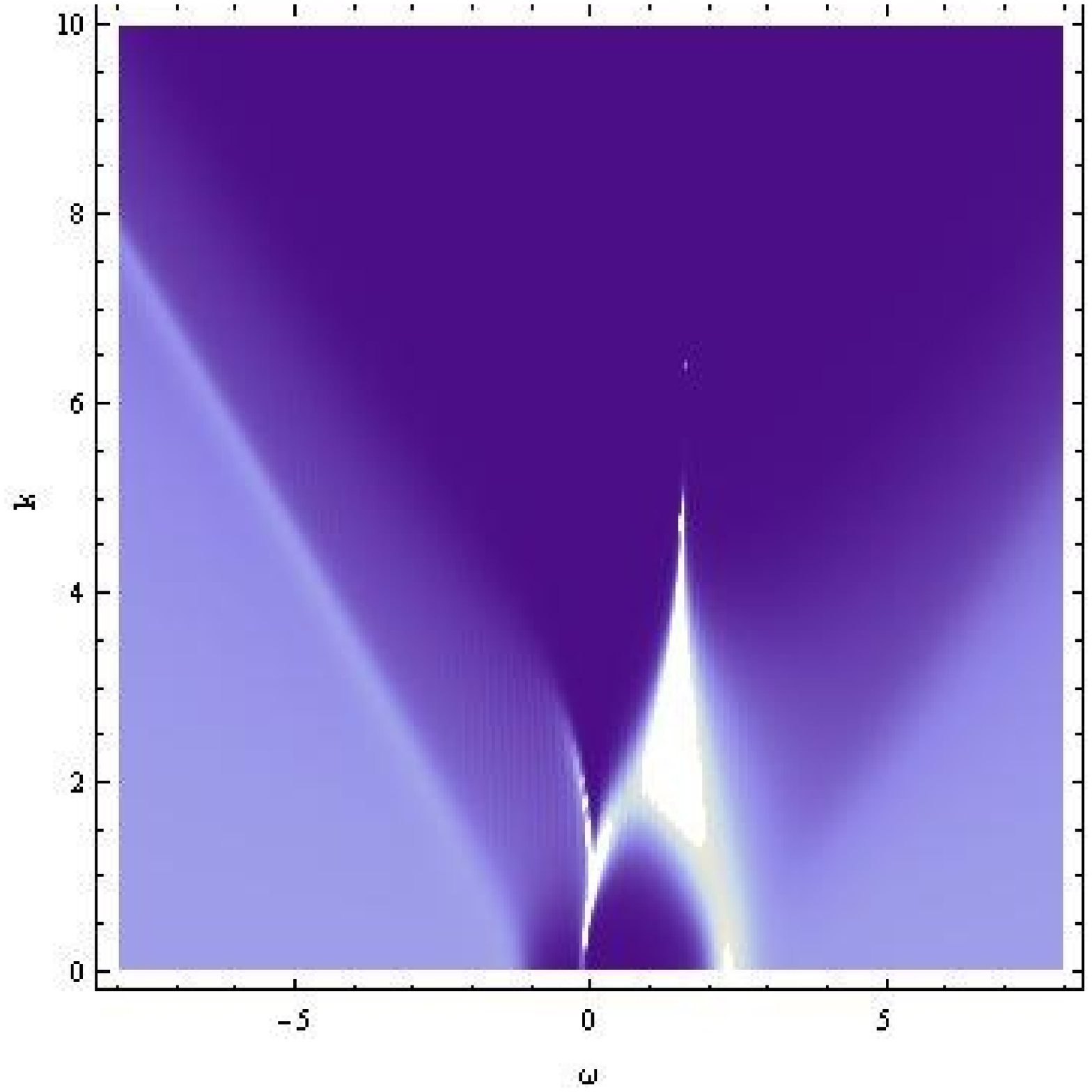}
\caption{The  plots of $A(\omega,k)$ for the case
of $q=1$ and  $p=2$.} \label{p2}
\end{figure}

\begin{figure}[ht]
\centering
\includegraphics[scale=0.3]{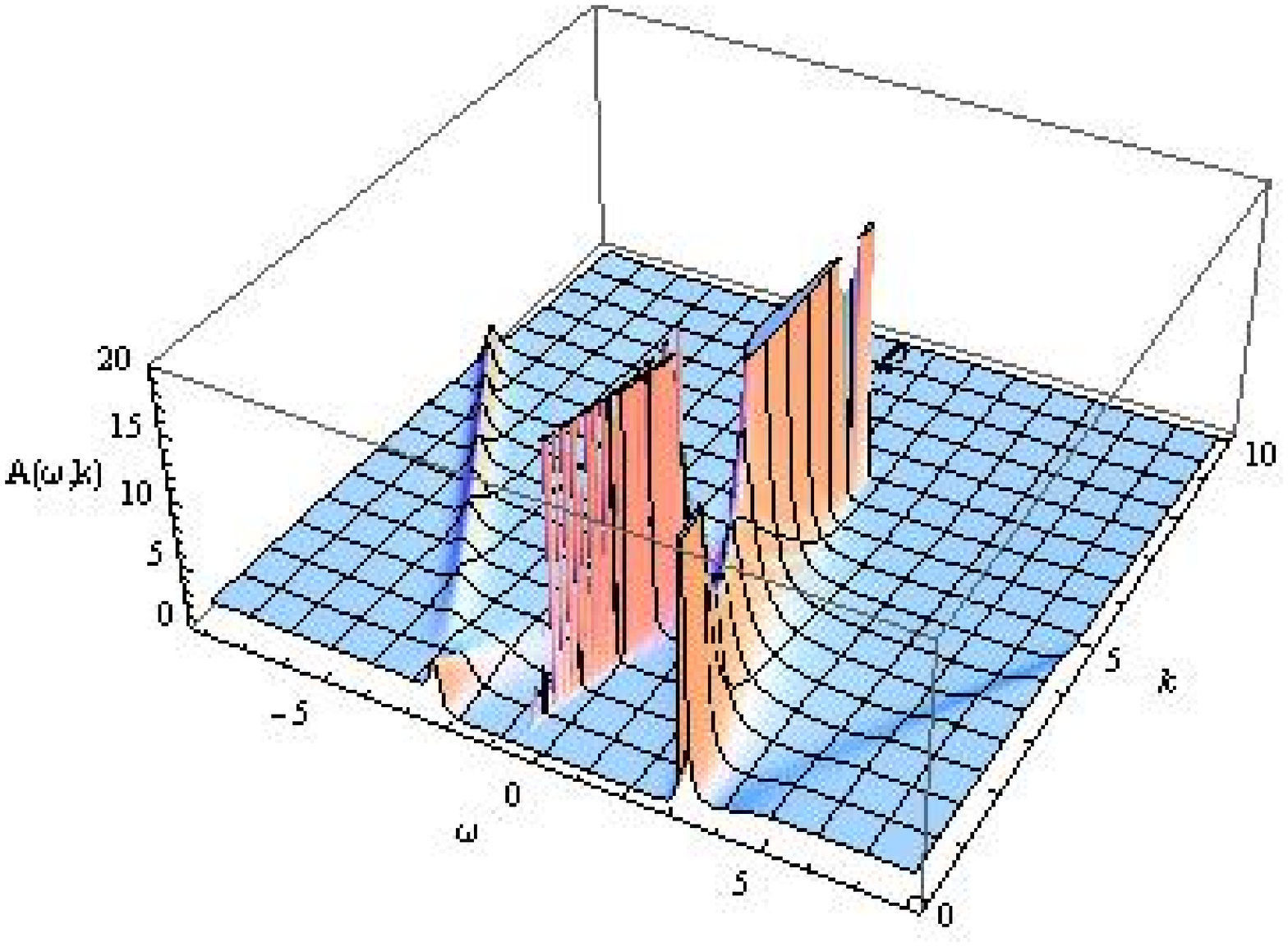}\hspace{1.5cm}
\includegraphics[scale=0.23]{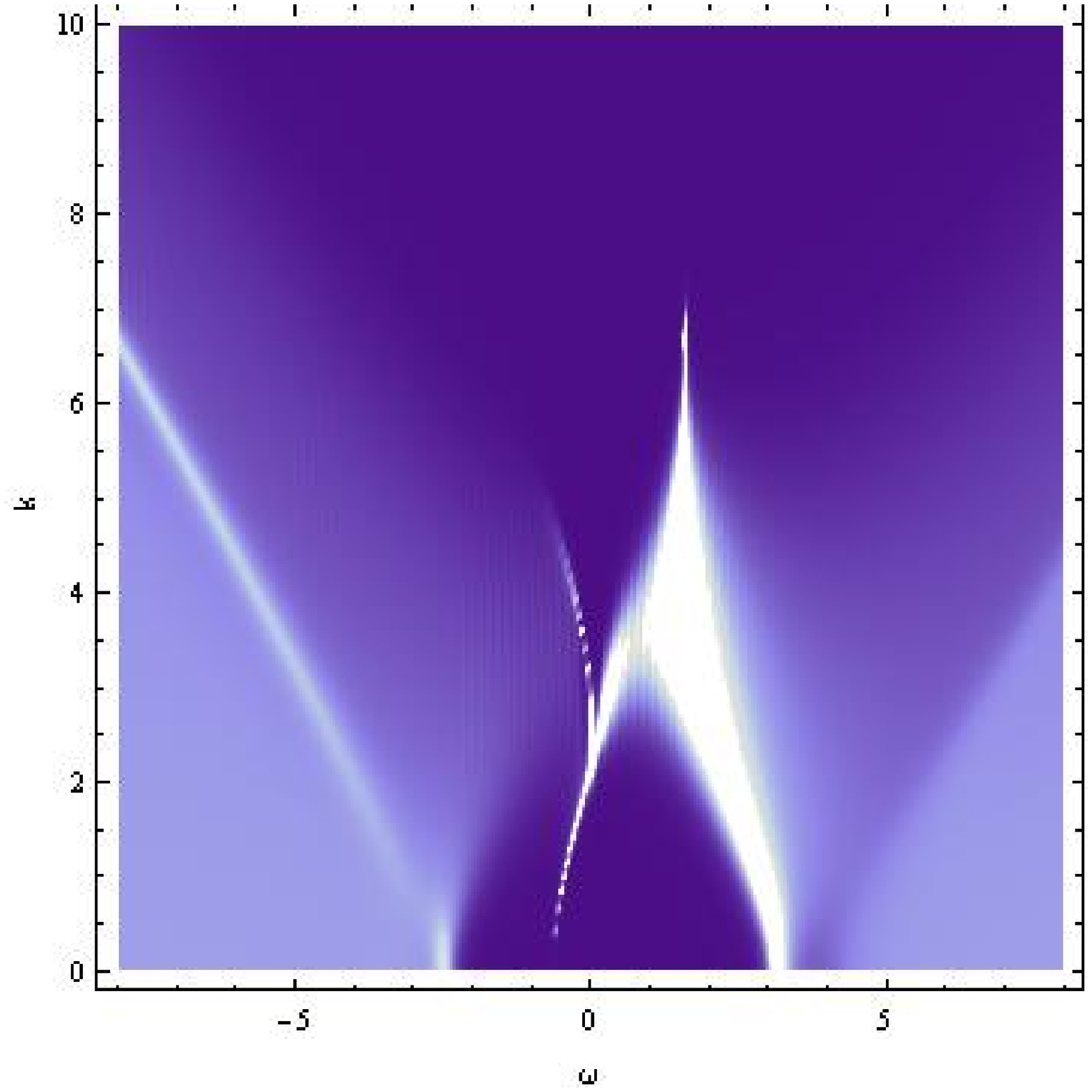}
\caption{The plots of $A(\omega,k)$ for the case
of $q=1$ and $p=4$.} \label{p4}
\end{figure}

\begin{figure}[ht]
\centering
\includegraphics[scale=0.3]{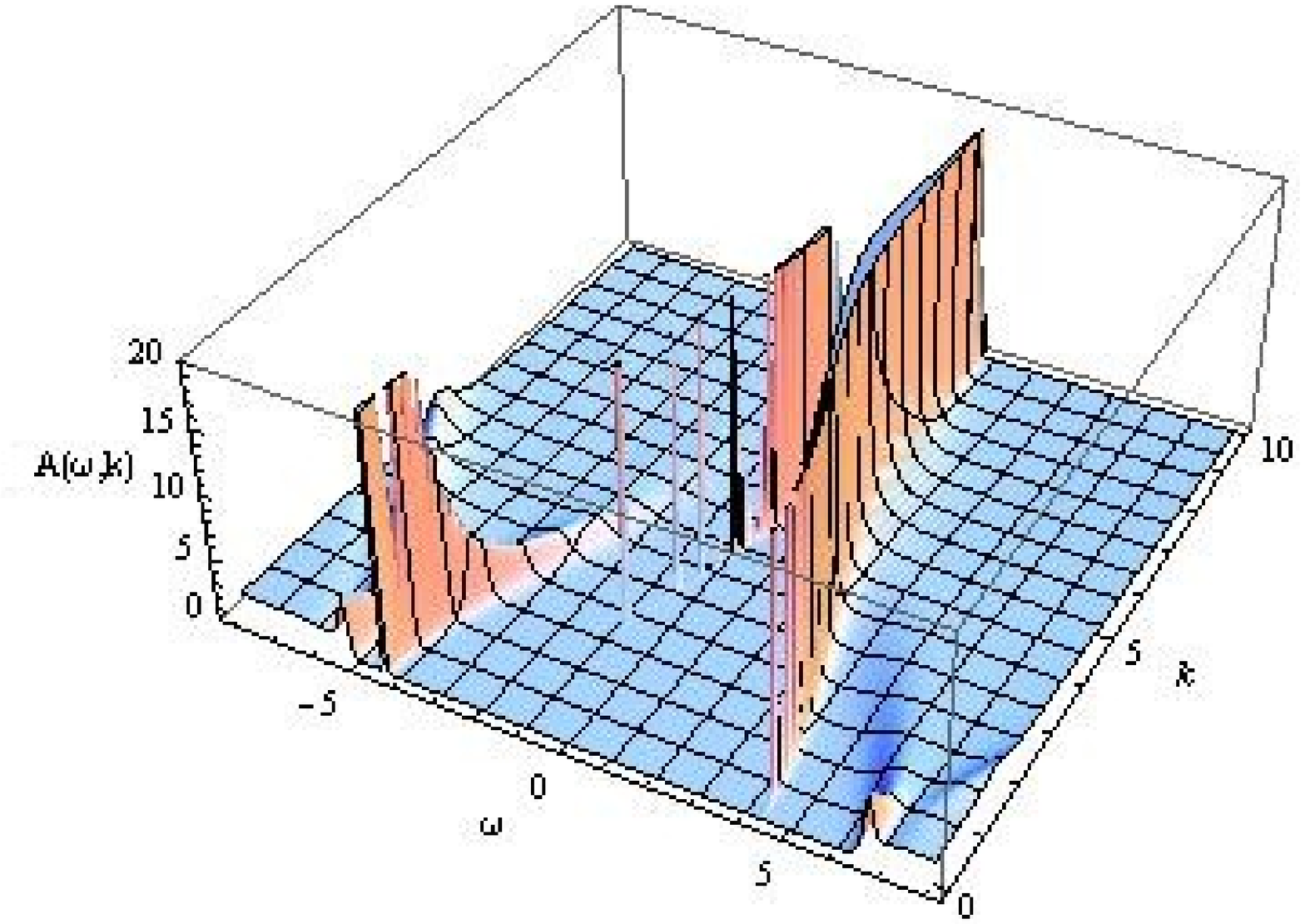}\hspace{1.5cm}
\includegraphics[scale=0.23]{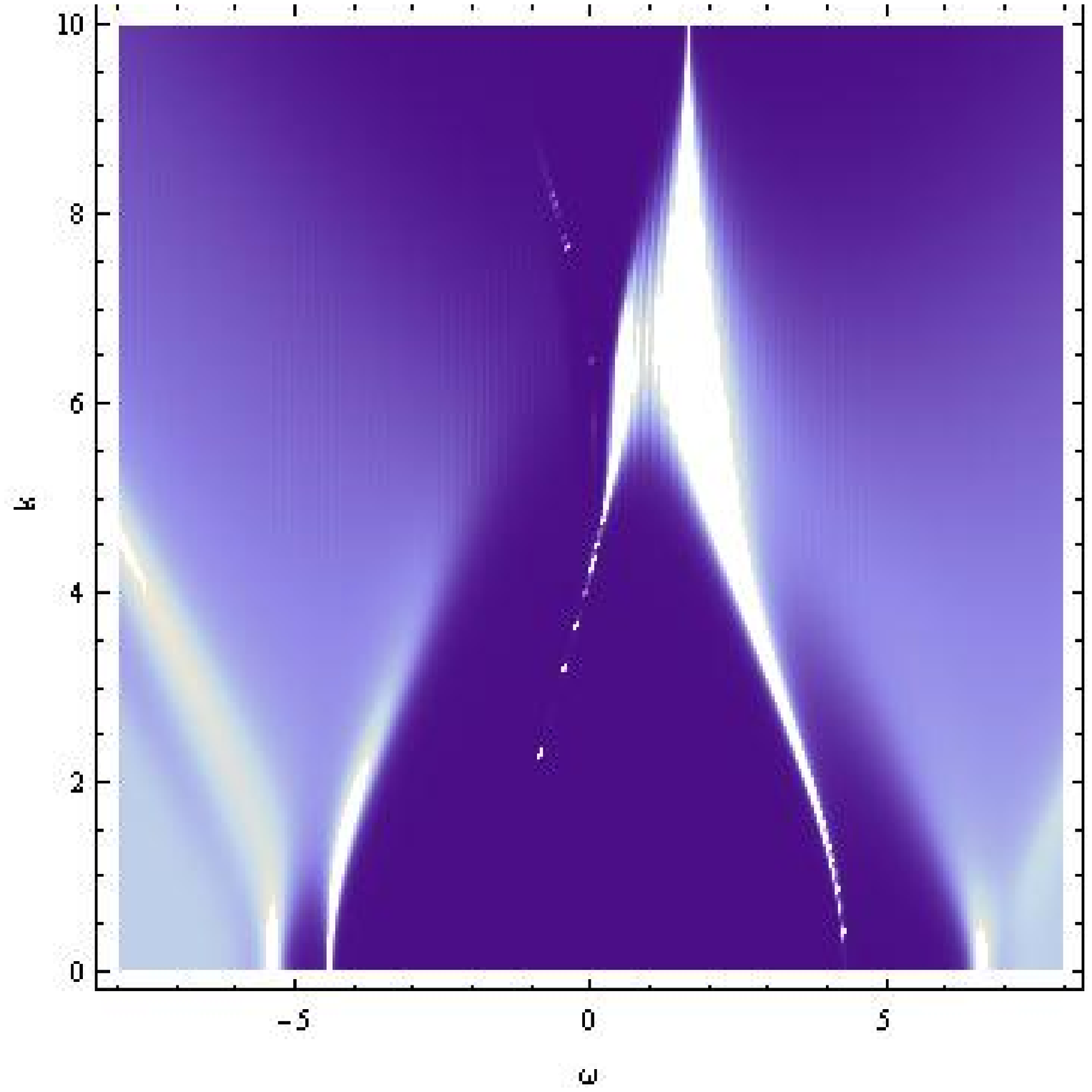}
\caption{The plots of $A(\omega,k)$ for the case
of $q=1$ and $p=8$.} \label{p8}
\end{figure}

We  choose $Q=1$ in this subsection. In our
computation  we first set $q=1$ . The numerical
results for the spectral function are shown in
Figs.\ref{p0}-\ref{p8}.

When there is no dipole coupling between the
fermonic field and the gauge field, $p=0$, the
result shown in Fig.\ref{p0} goes back to figure
1 in \cite{WJL2} for the non-relativistic fixed
point. Clearly, we see that $A(\omega,k)$ shows
the flat band with the peculiar property of poles
distributing continuously at a finite interval of
momenta. This finite band is mildly dispersed at
low momentum, but it presents strong peak at high
momentum. This is because the high momentum
modes sit outside the lightcone and can't decay.
The peak tends sharper as can be checked by
plotting a suitable cross section at fixed high
$k$.

We now turn on the dipole coupling parameter $p$
and report the results  in
Figs.\ref{p2}-\ref{p8}. At low momentum, the flat
band gets more dispersed and the band recovers
flatness at higher momentum when the non-minimum
fermion coupling becomes stronger.   With the
increase of the dipole coupling, the frequency of
the flat band is larger and the peak of the flat
band becomes sharper at small momenta, which
means that more energy is needed to excite the
small momentum modes for larger $p$. A finite gap
opens up and enlarges  when the dipole coupling
becomes strong. The dipole coupling $p$ drives
the dynamical formation of a gap, which mimics
the role of interaction strength $U/t$ in the
Hubbard model, where the Mott gap forms once
$U/t$ exceeds a critical value. The further
extreme behavior of the flat band shows the
physics approaching a more strongly coupled
corner when the dipole coupling becomes stronger.
Furthermore we observe that at large enough
momentum the flat band always shifts to
$\omega=\sqrt{3}$ for all dipole coupling. This
is because the frequency $\omega$ is measured
relative to the chemical potential. The flat band
corresponds to some zero modes in the Minkowski
vaccum, with the vanishing absolute energy of
Fermion  characterized by $\omega_{eff}=\omega+q
A_t$ in the Dirac equation
(\ref{DiracEinFourier}) on the boundary. From
(\ref{blackhole})(\ref{tmu}), it is clear that $A_t=\mu$ at the
boundary so that the frequency $\omega$ relates
to the chemical potential at the boundary, which
is independent of the dipole coupling.

The appearance of the flat band for the
non-relativistic fermions observed here is
interesting, which has not been observed for the
relativistic fermions in the charged dilatonic
AdS black hole background \cite{Wen}. The
different properties between the non-relativistic
and relativistic fermions observed here support
the findings in \cite{WJL1} by comparing with
\cite{R.G.Leigh1} for the charged AdS black hole
background.
\begin{figure}[ht]
\centering
\includegraphics[scale=0.3]{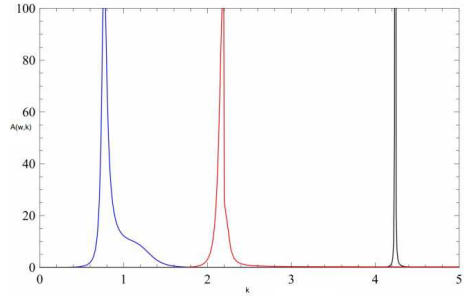}\hspace{1.5cm}
\caption{The function of $A(\omega,k)$ at
$\omega=-10^{-7}$ for different dipole coupling.
 The lines from left
to right are for $p=2,4$ and $8$, respectively.
In the computation we take $q=1$.} \label{kfq1}
\end{figure}
\begin{table}
\centering
\begin{tabular}{|c|c|c|c|c|}
  \hline
  $p$ & 0 &2 & 4 &8   \\ \hline
  $k_F$ &No & 0.76806283& 2.19010221 & 4.23765084 \\ \hline
  $v_F$ &No & 0.286127& 0.444296 & 0.401080 \\ \hline
\end{tabular}
\caption{The Fermi momentum and Fermi velocity
for different dipole couplings when we choose
$q=1$.} \label{tablekfq1}
\end{table}
\begin{figure}[ht]
\centering
\includegraphics[scale=0.48]{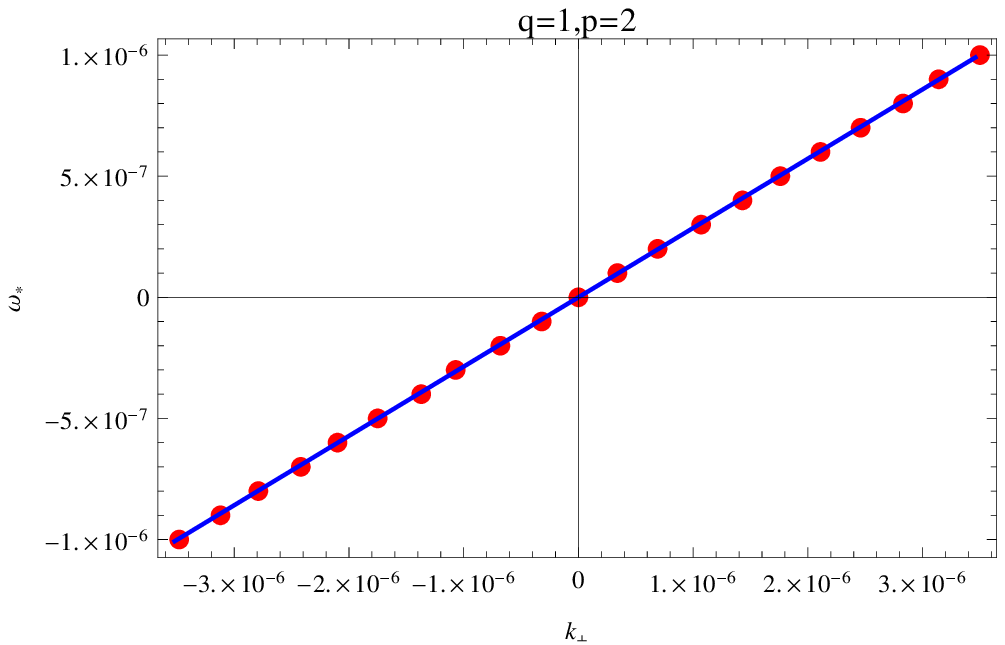}\hspace{0.5cm}
\includegraphics[scale=0.48]{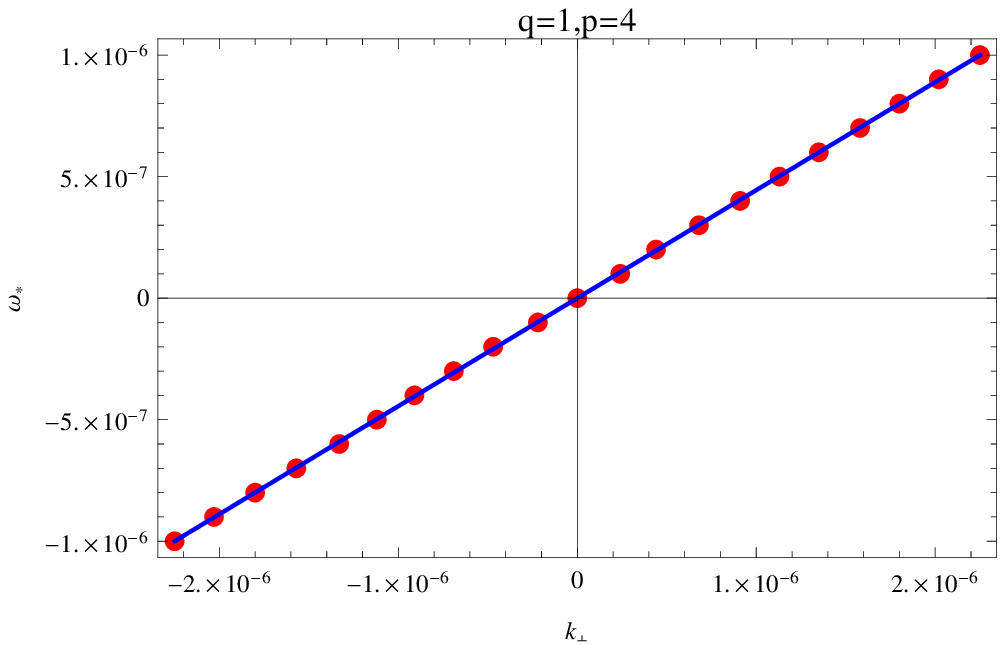}\hspace{0.5cm}
\includegraphics[scale=0.48]{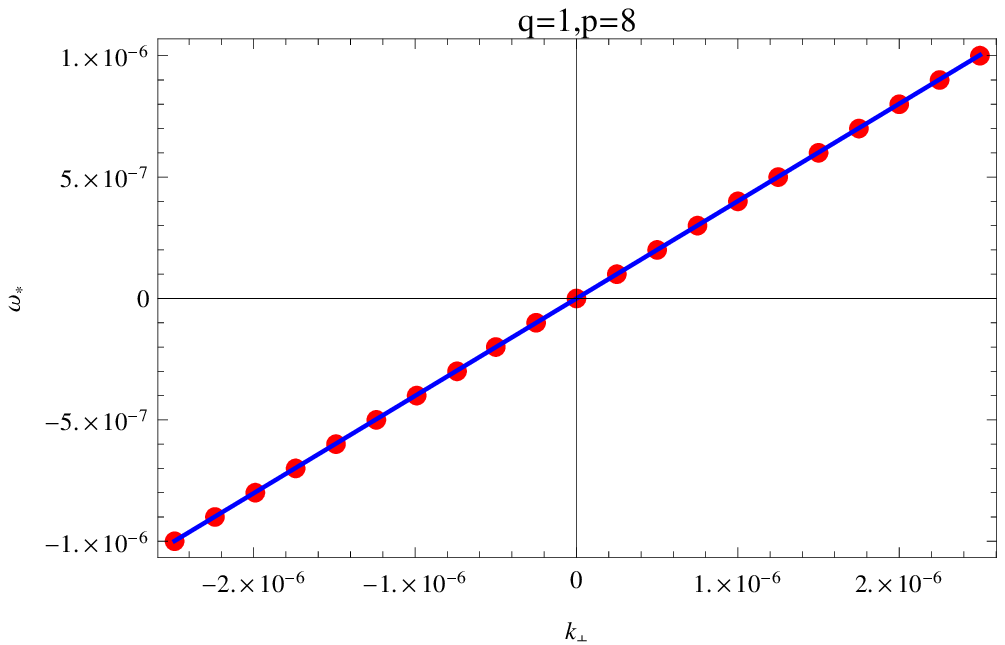}
\caption{The dispersion relation near the Fermi
surface at the Non-relativistic fixed point of
$q=1$ for various $p$.} \label{kwq1}
\end{figure}

Now we turn to discuss the fermi momentum. From
Fig.\ref{p0}-\ref{p8} we see  that the Fermi
surface opens up for nonzero $p$. In the limit
$\omega \rightarrow 0$, the sharp peak of the
spectral function represents the fermi surface.
For various values of $p$ we illustrate the
location of Fermi surface in Fig.\ref{kfq1}. We
find that the corresponding Fermi momentum
increases when the dipole interaction becomes
stronger as shown  in Table.\ref{tablekfq1}. We
can further investigate the dispersion relation
near the Fermi surface. For various $p$ we show
in Fig.\ref{kwq1}  that the linear relation
between $\omega$ and $k_{\bot}=k-k_F$ behaves as
\begin{equation}
\omega \simeq v_F k_{\bot}.
\end{equation}
This linear dispersion relation indicates that
the excitation near the Fermi surface is
well-defined and the Fermi liquid is like the
Landau Fermi liquid. This property is not
influenced by the dipole coupling, which is
consistent with the case of relativistic fermions
discussed in \cite{Wen}. The values of the Fermi
velocity $v_F=\frac{\partial \omega}{\partial k}$
for different dipole coupling are listed in
Table.\ref{tablekfq1}.
\begin{figure}[ht]
\centering
\includegraphics[scale=0.3]{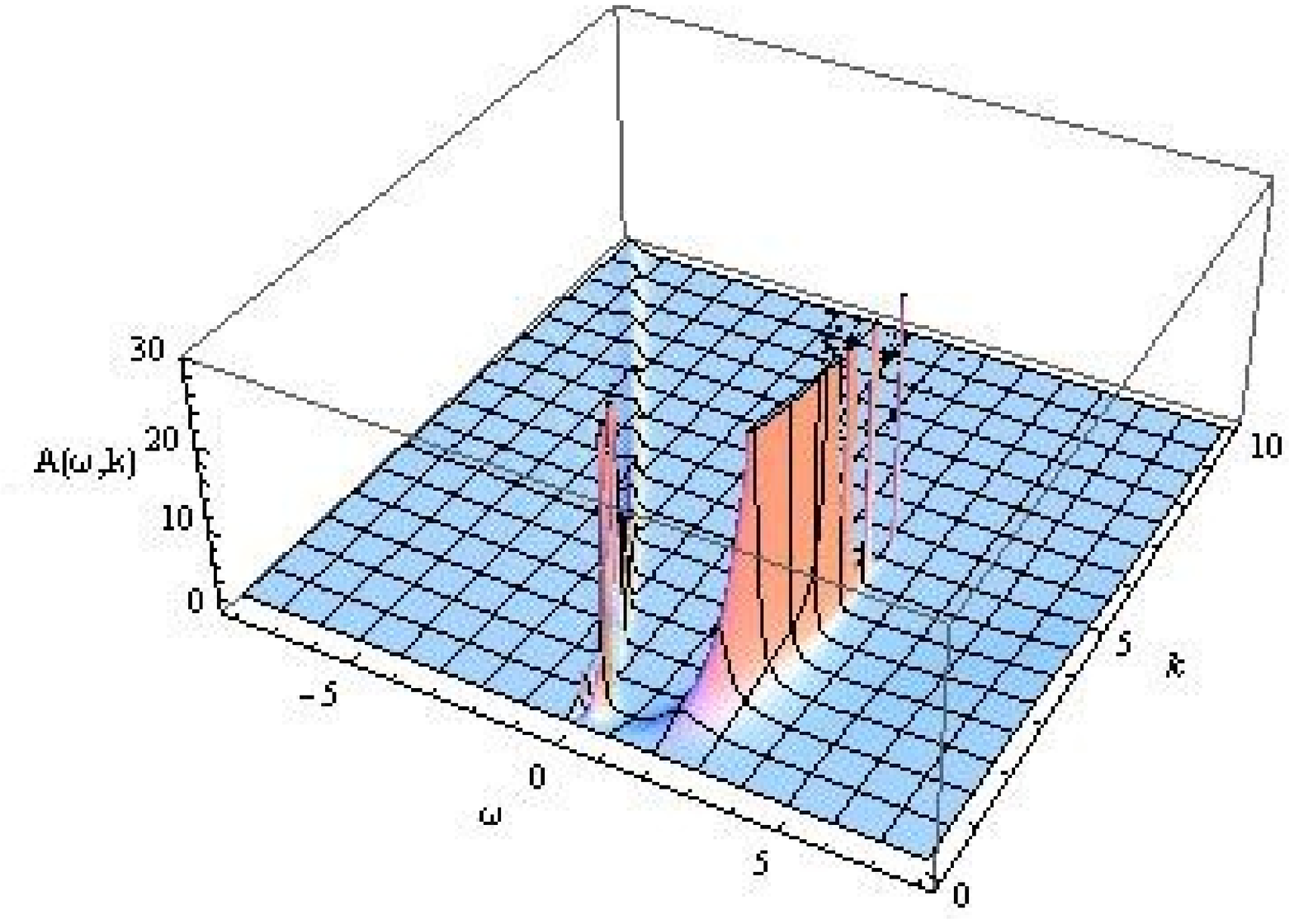}\hspace{1.5cm}
\includegraphics[scale=0.23]{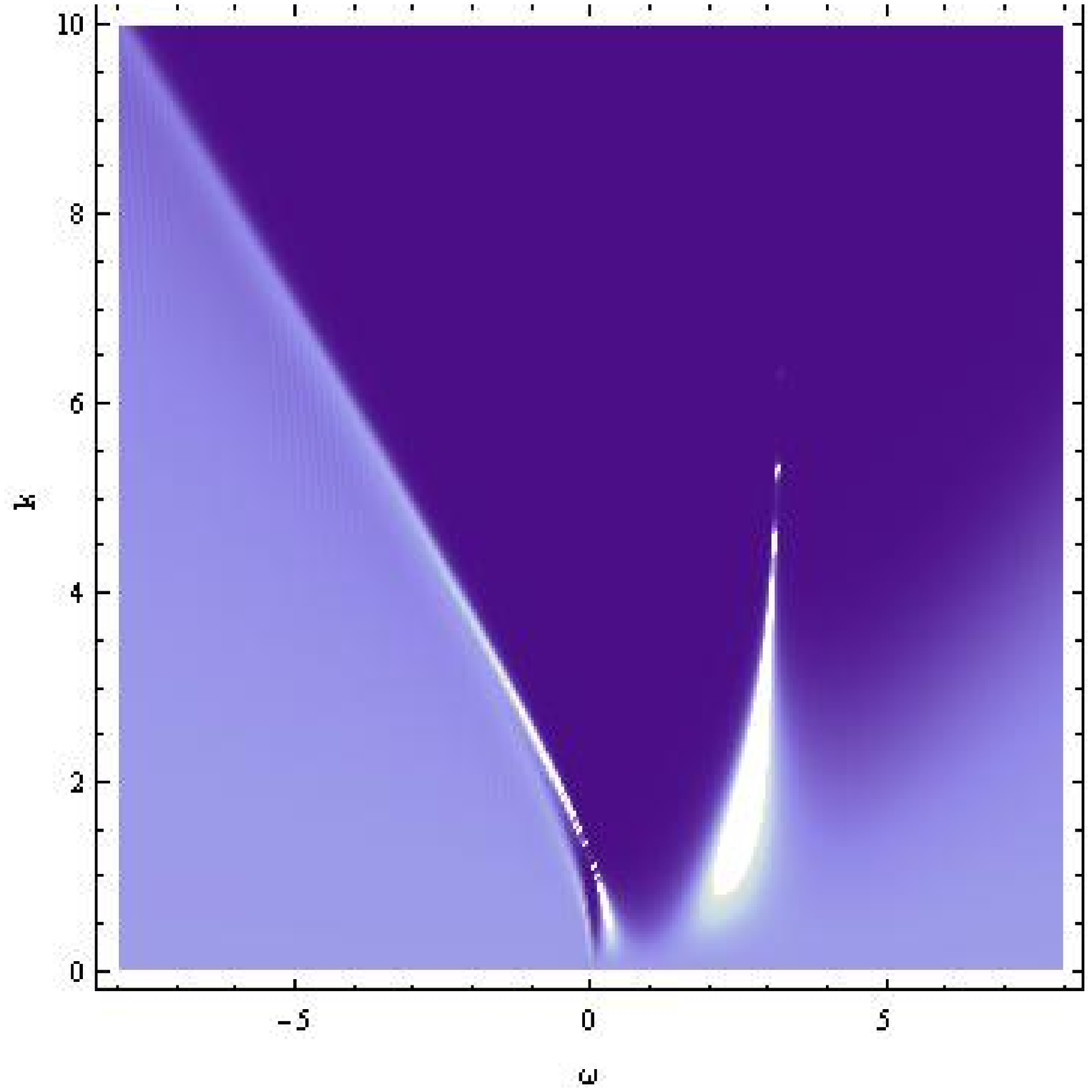}
\caption{The plots of $A(\omega,k)$ for the case
of $q=2$ and $p=0$.} \label{q2p0}
\end{figure}
\begin{figure}[ht]
\centering
\includegraphics[scale=0.3]{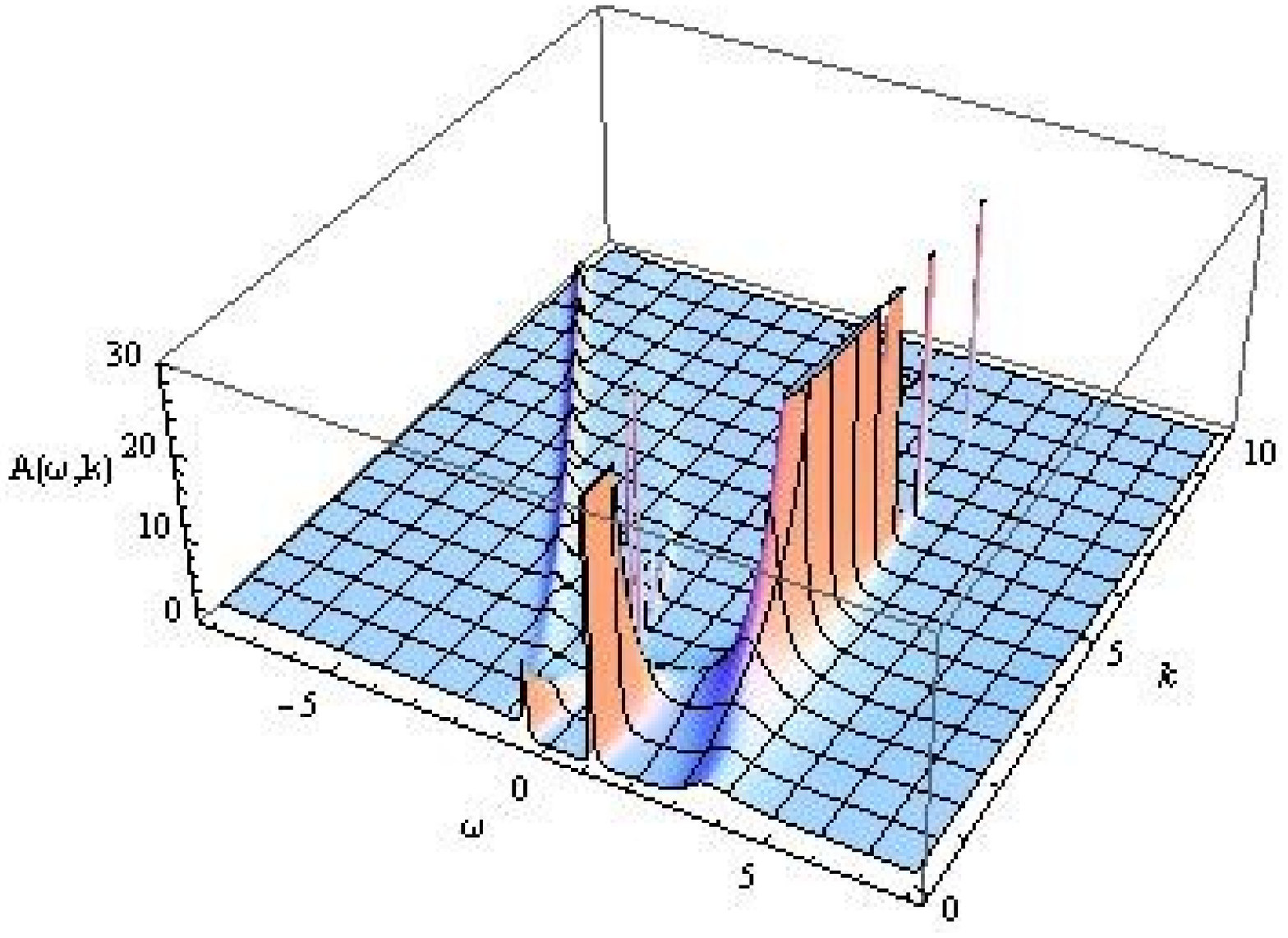}\hspace{1.5cm}
\includegraphics[scale=0.23]{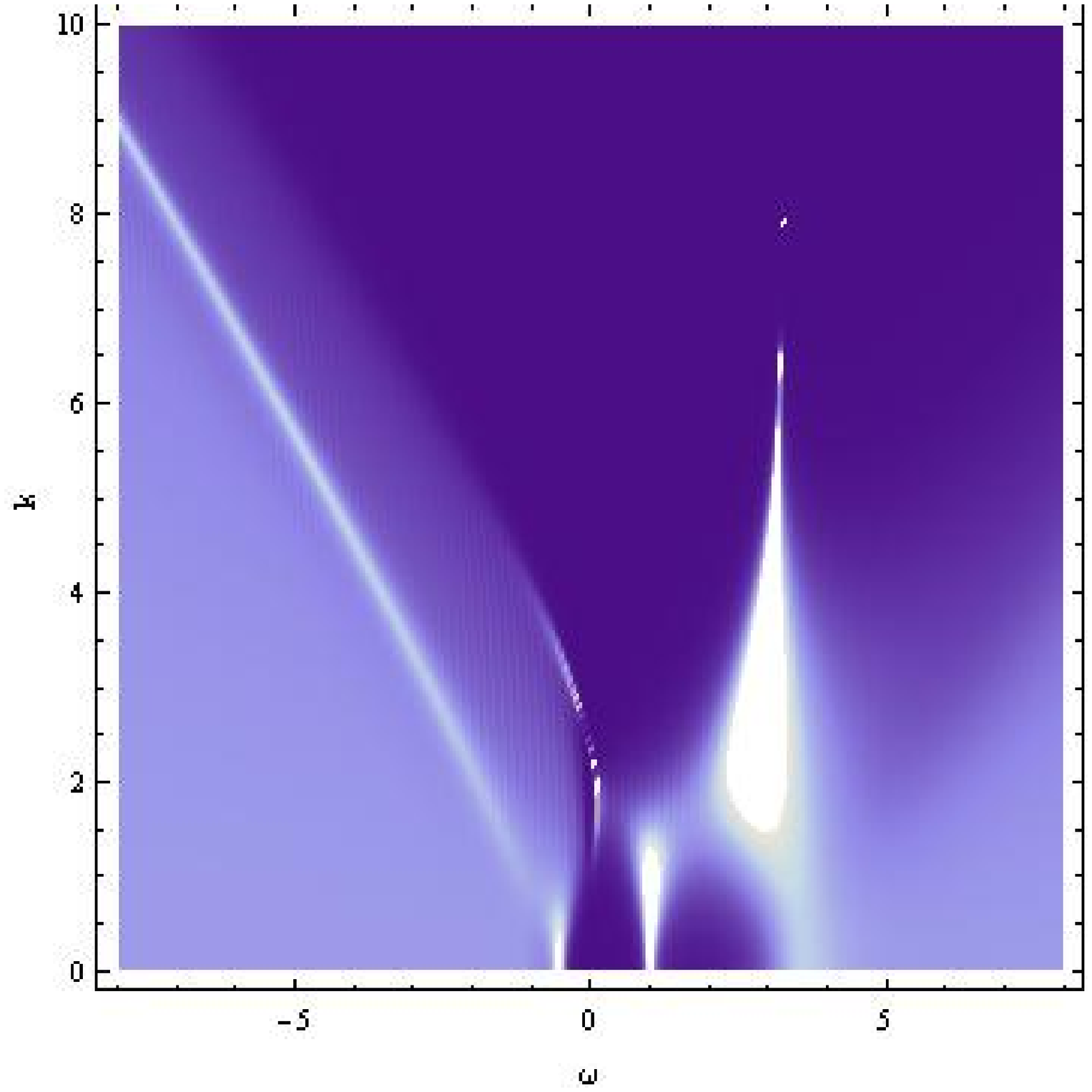}
\caption{The plots of $A(\omega,k)$ for the case
of $q=2$ and $p=2$.} \label{q2p2}
\end{figure}
\begin{figure}[ht]
\centering
\includegraphics[scale=0.3]{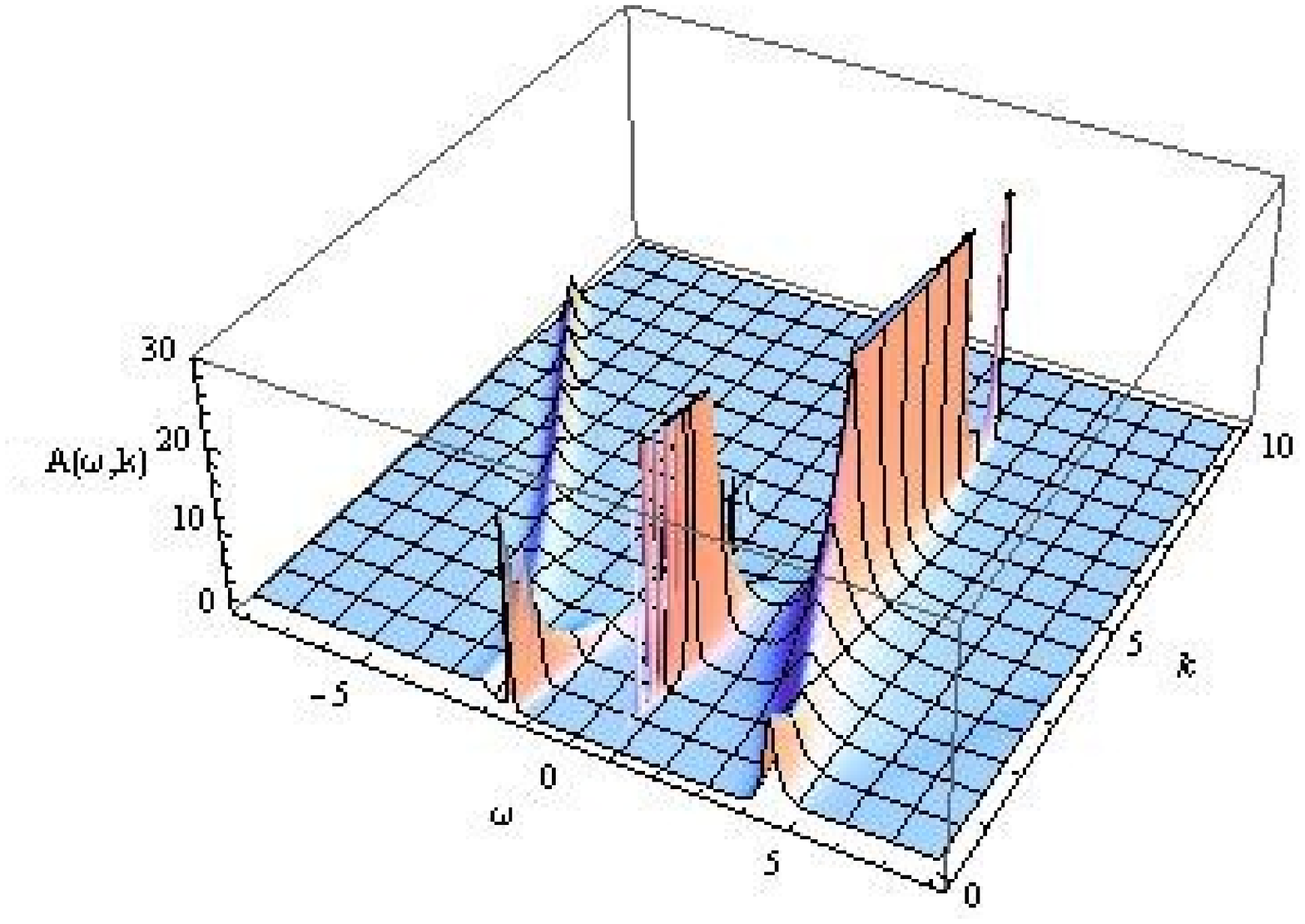}\hspace{1.5cm}
\includegraphics[scale=0.23]{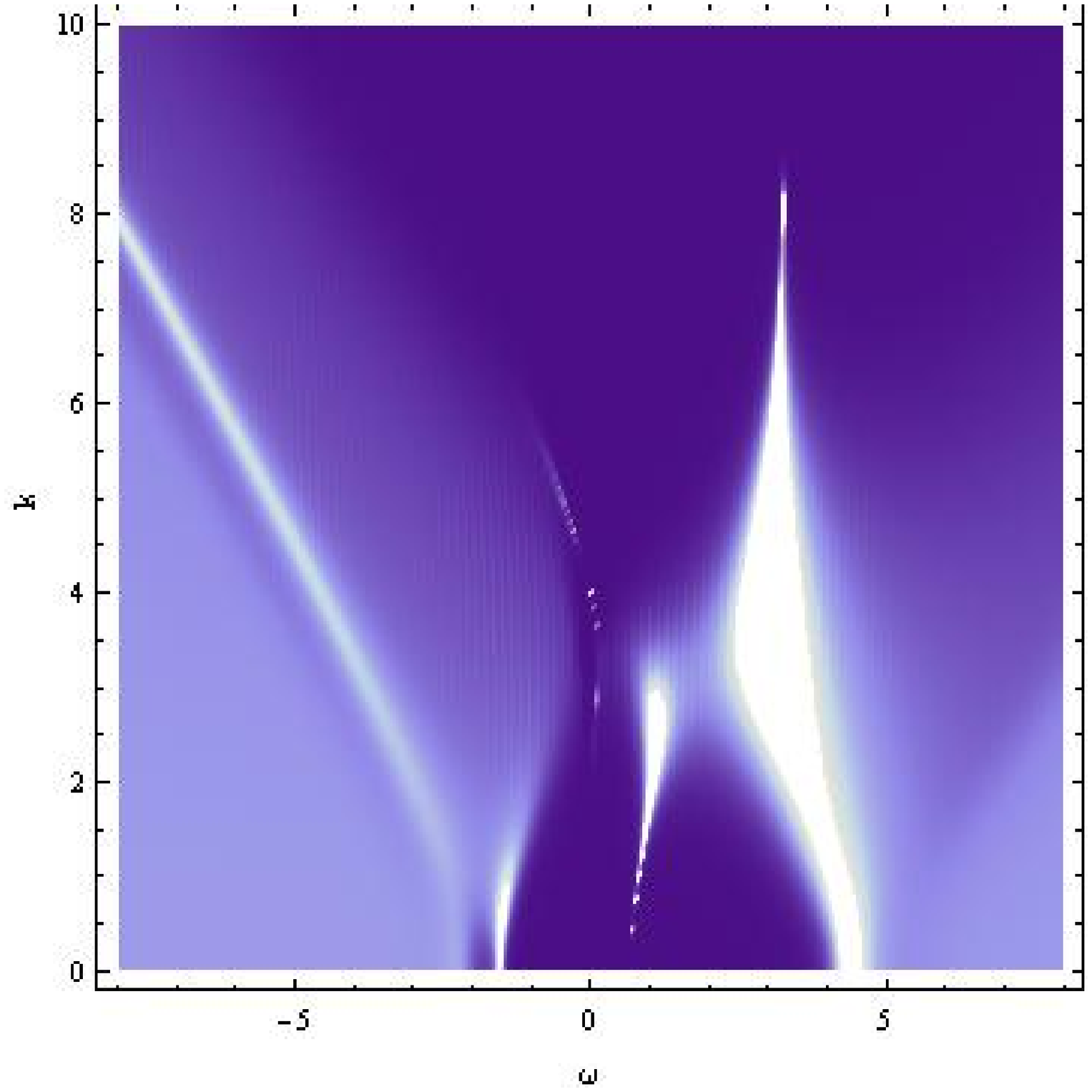}
\caption{The plots of $A(\omega,k)$ for the case
of $q=2$ and $p=4$.} \label{q2p4}
\end{figure}
\begin{figure}[ht]
\centering
\includegraphics[scale=0.3]{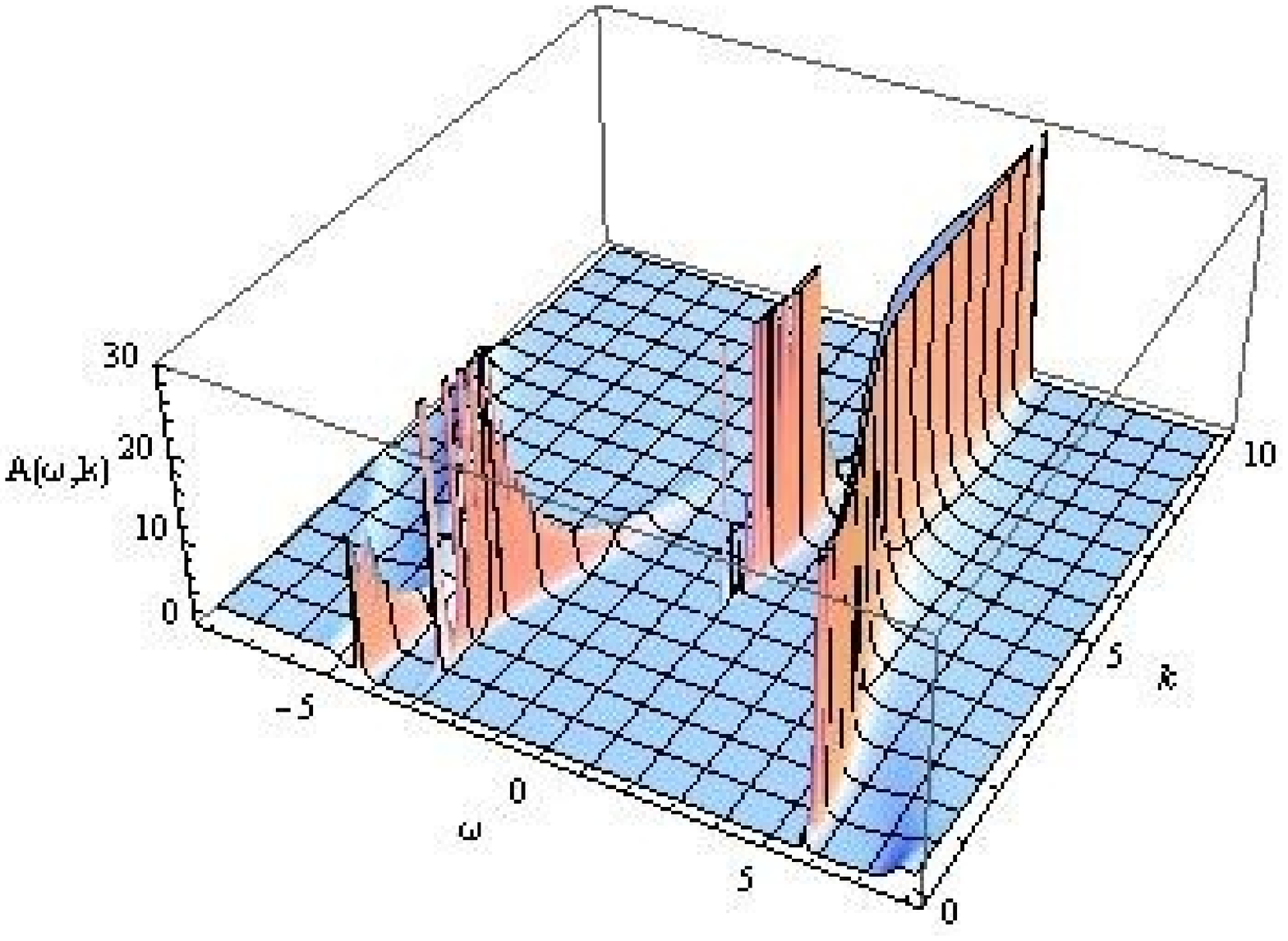}\hspace{1.5cm}
\includegraphics[scale=0.23]{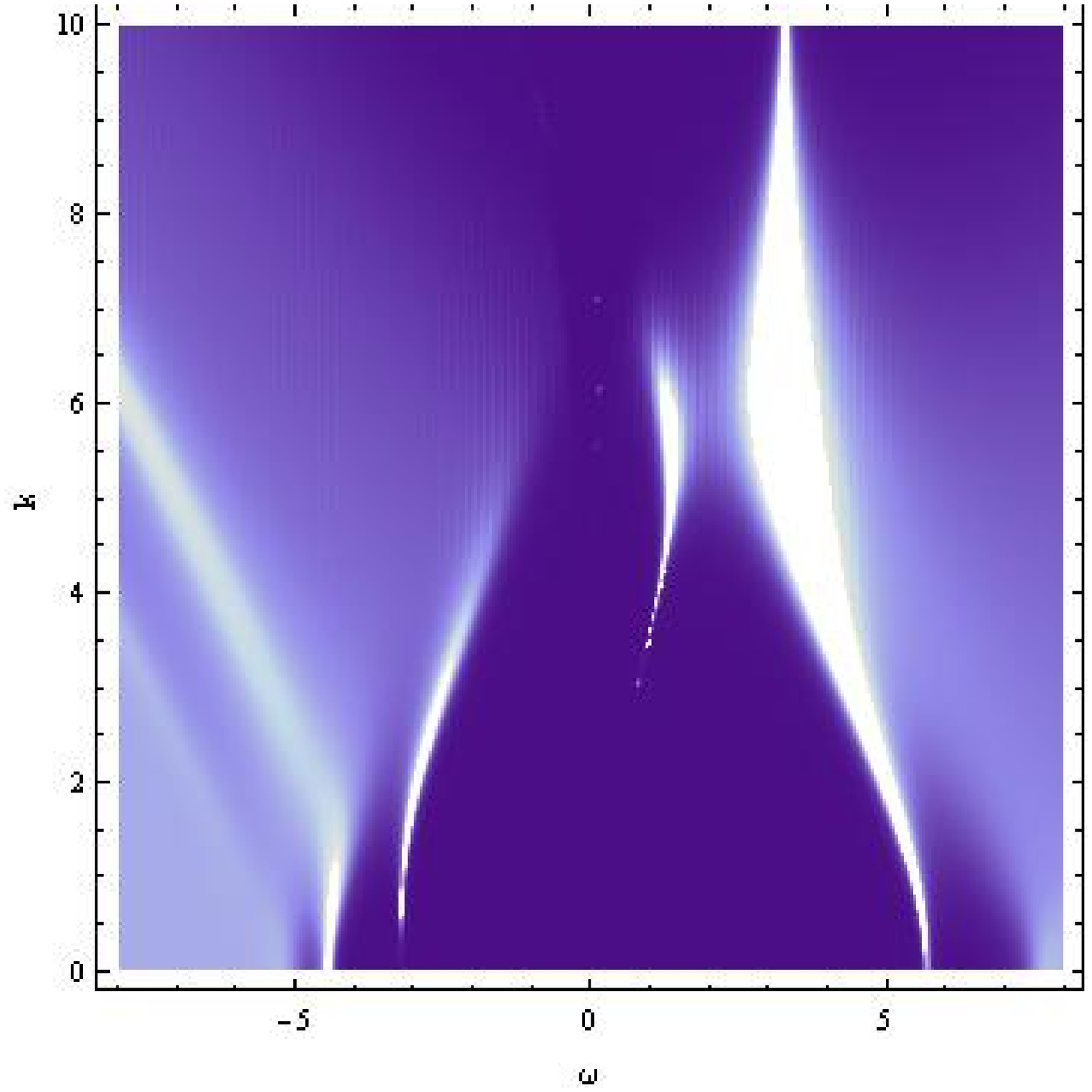}
\caption{The plots of $A(\omega,k)$ for the case
of $q=2$ and $p=8$.} \label{q2p8}
\end{figure}

As we discussed above that the bulk dipole
coupling indeed has imprint on the holographic
system. It is natural to ask its role in the
properties of the fermion operator in the dual
theory. In the vacuum, the coupling does not
affect the fermion-fermion Green function.
However, it will influence the algebraic
structure of the current-fermion-fermion three
point correlation function. Though the full story
of the dipole effect from CFT is complicated, one
simple character  we can see from (\ref{actionspinor}) is that
when $p=0$, there are no terms in
$\langle\bar{\zeta_{\alpha}}J^{\mu}\zeta_{\beta}\rangle$
proportional to the second rank clifford algebra
element $\Gamma^{\mu\nu}$, while for nonzero
dipole couplings, such terms will
arise\cite{McGreevy}. As the dipole coupling
increases, this makes the Fermi surface pole fall
into a log-oscillatory
region\cite{HongLiuNonFermi}, disappear and
finally a gap forms. It would be interesting to
analytically understand why the dipole coupling
change the shape of the Fermi surface and the
oscillatory region and further explore the full
story in the field theory. This will serve as a
separate work in the future.

\subsubsection{The influence of fermion charge and the dilaton field}

The chemical potential felt by the probe fermion
is read as $\Omega=q\mu$. Since $g_F$, the
effective dimensionless gauge coupling, is
implied in $\mu$, we have $\Omega\sim g_{F}q$. In
the conformal field theory, the fermion charge
$q$ or the gauge coupling $g_{F}$ denotes the
interaction strength between the spinor and
massless vector fields. It will modify the
$n-$point function for the boundary spinors by
bringing the vector-spinor-spinor function. The
detailed calculation from the AdS/CFT
correspondence with this interaction was shown in
\cite{Muck}. The authors of
\cite{HongLiuNonFermi} have reported that the
properties of the holographic fermion are charge
dependent. On the other hand, here from the bulk
action, we can see that the dilaton field is
related to the gauge coupling $g_F$ through the
form $g_{F}^{2}=4e^{-\phi}$. This shows that the
dilaton field affects the gauge coupling and in
turn will have imprint on the chemical potential
of the boundary field theory. Thus, it is of
interest to study how the fermion charge and
dilaton field affect the spectral function,
respectively.

In \cite{J.Ren}, it
was claimed that the charge $q$ can affect the
excitation near the Fermi surface in charged dilaton black hole.
In addition, it was argued that there is some kind of
competition between the charge $q$ and the dipole
coupling $p$ to create the Fermi surface for the
non-relativistic fixed point in charged AdS black
hole \cite{WJL2}. Here we would like to further
examine the effect of $q$ in the Fermi system in
the charged dilatonic AdS background. We will
change the value of $q$ in the computation and
compare  with the result for $q=1$. The spectral
functions are shown in Fig.\ref{q2p0}-\ref{q2p8}
for taking $q=2$. Similar to the results by
setting $q=1$, flat bands tend to $2\sqrt{3}$ at
large enough momentum which is independent of the
dipole coupling. Besides the similarity, we
observe the different property for choosing
different $q$. When $q=2$, we see that the
quasi-particle like peak, i.e., Fermi surface
appears even at $p=0$. This is different from
small $q$ case where there is no Fermi surface in
the $\omega=0$ limit for the minimal Fermion
coupling. Moreover, the emergence of the gap can
happen at smaller dipole coupling when $q$ is
bigger, for example the gap emerges around $p\sim13$
for $q=1$, while the gap appears around
$p\sim8.5$ when $q=2$. These observations support
that $q$ is nontrivial, it influences the
Fermi system.
\begin{table}
\centering
\begin{tabular}{|c|c|c|c|c|}
  \hline
  $p$ & 0 &2 & 4 &8   \\ \hline
  $k_F$ &1.18628543 & 2.33984327& 3.99854464 & 7.40908575 \\ \hline
  $v_F$ &-0.607729  & -0.380546 & -0.380714 &-0.382852  \\ \hline
\end{tabular}
\caption{The Fermi momentum and Fermi velocity
for different dipole couplings for $q=2$.}
\label{tablekfq2}
\end{table}

The Fermi momentum and Fermi velocity for
choosing $q=2$ are listed in
Table.\ref{tablekfq2}. Similar to setting $q=1$,
the Fermi momentum increases with $p$  and the
dispersion relation is kept linear. We see that
the dispersion relation is always linear and is
independent of the fermion charge and the dipole
coupling in this case. Comparing with the
relativistic situation \cite{Wen}, we find that
for the same bulk coupling, the Fermi momentum is
smaller in the non-relativistic case. The
suppression of the Fermi momentum in the
non-relativistic case can attribute to the
presence of the flat band. In addition, we find
that the Fermi velocities for $q=2$ and $q=1$
have opposite signs. These imply that the charge
$q$ do influence the excitation near the Fermi
surface which supports the claim in \cite{J.Ren}.
\begin{figure}[ht]
\centering
\includegraphics[scale=0.25]{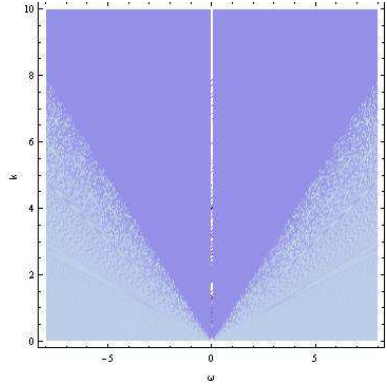}\hspace{1.5cm}
\caption{$A(\omega,k)$ with $q=p=2$ for $Q=0$.}
\label{figq1Q0p}
\end{figure}
\begin{figure}[ht]
\centering
\includegraphics[scale=0.23]{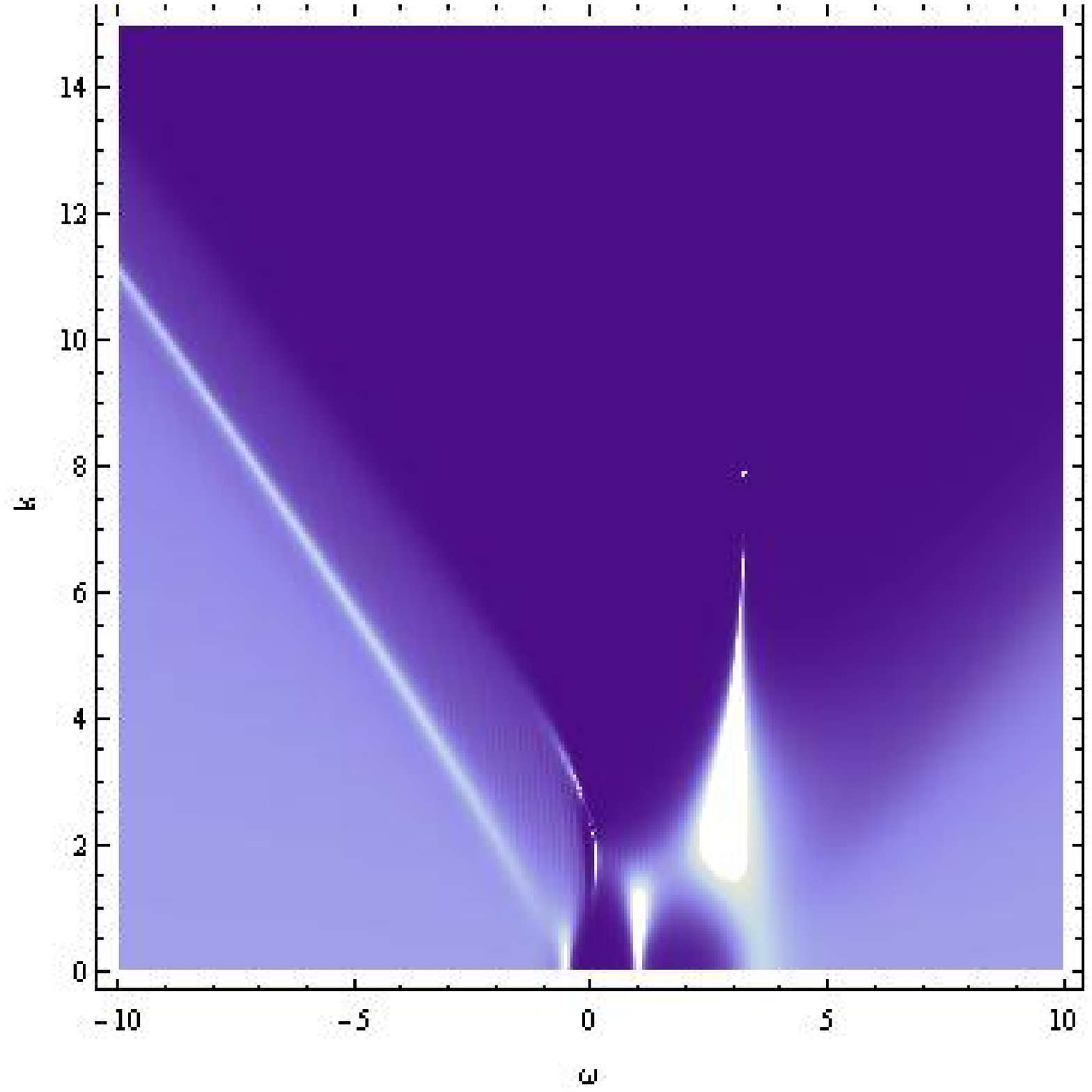}\hspace{1cm}
\includegraphics[scale=0.23]{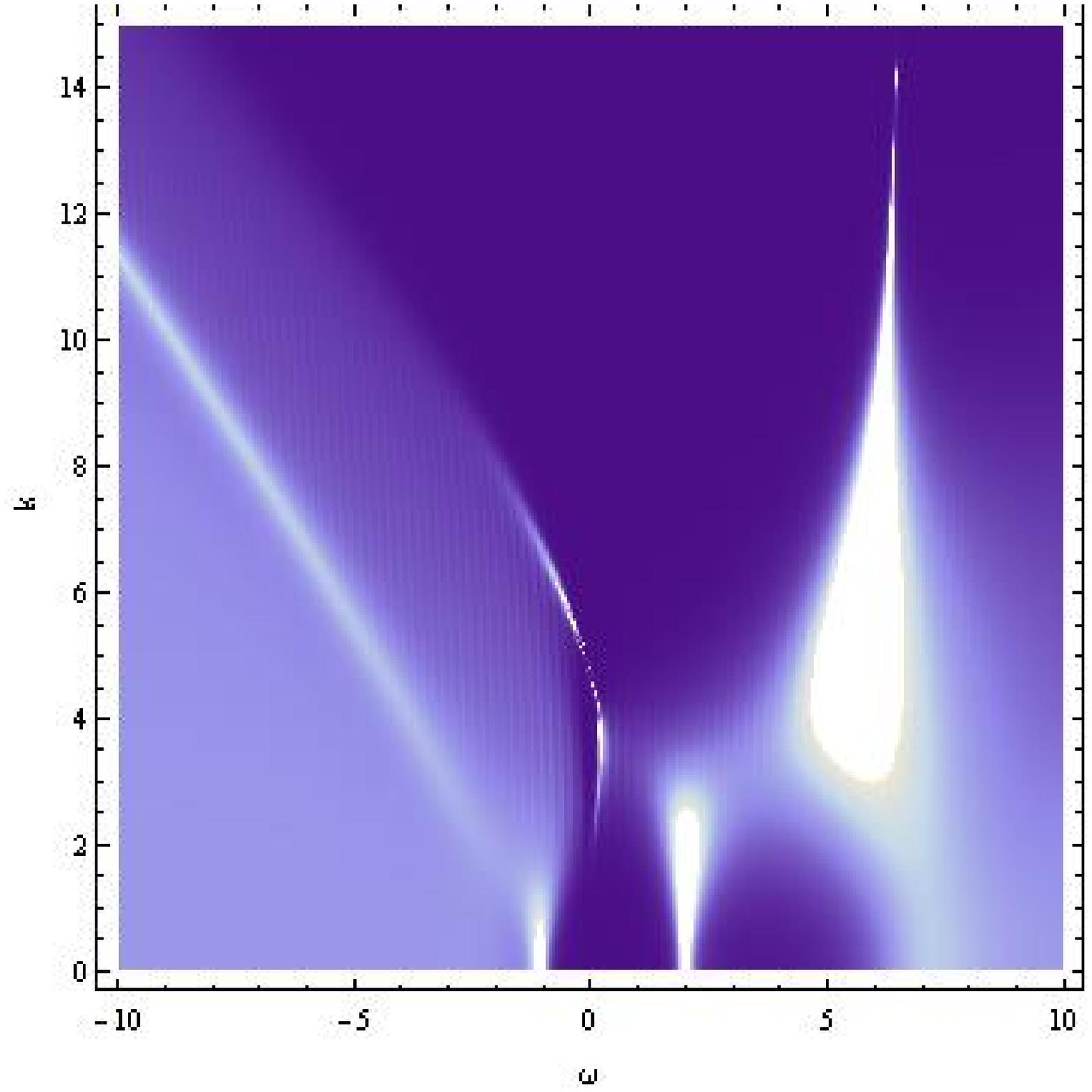}\hspace{1cm}
\includegraphics[scale=0.23]{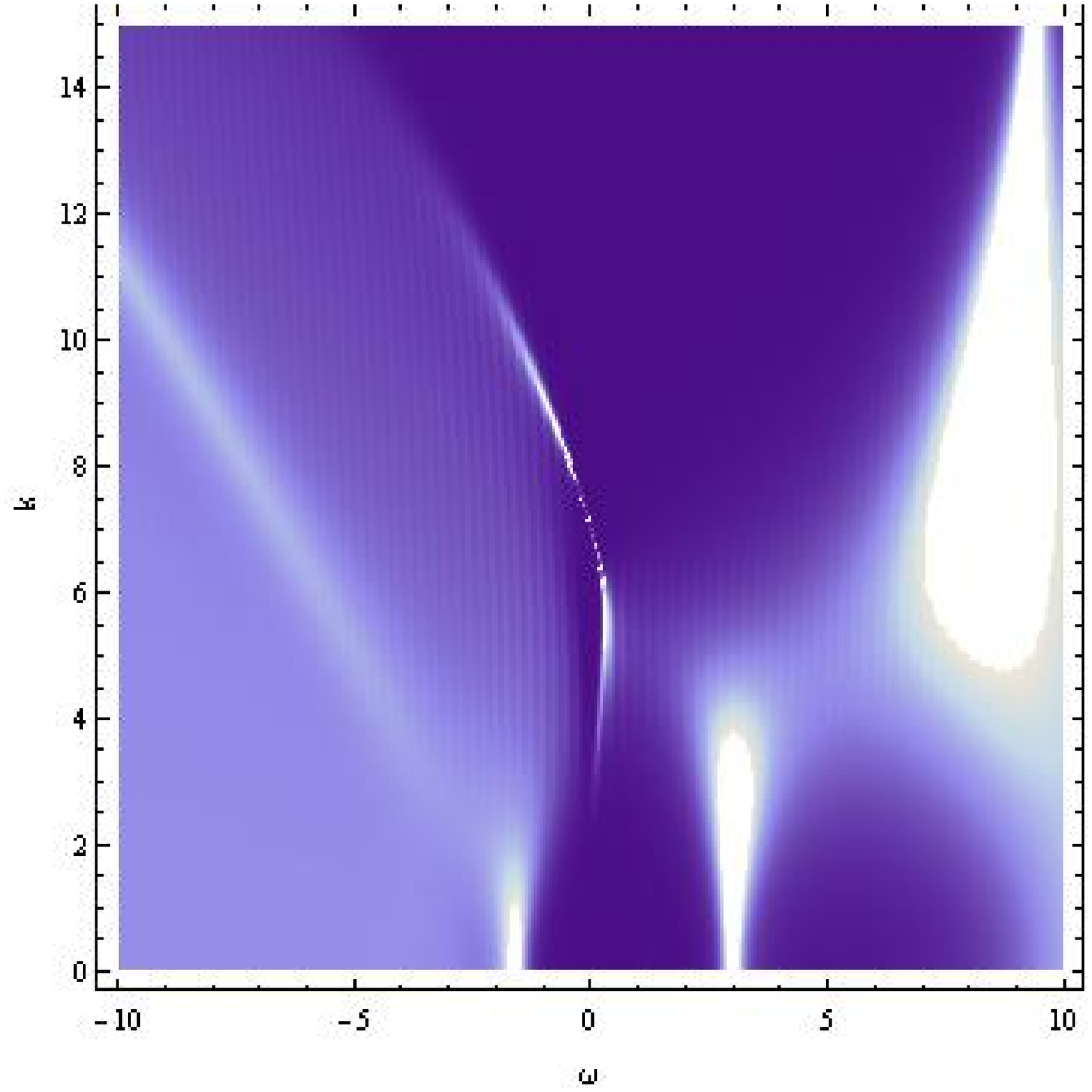}\hspace{1cm}
\caption{The plots of $A(\omega,k)$ with $q=p=2$. Here $Q$ is
set to be $1, 2$ and $3$ respectively from left
to right. } \label{densitydiffQ}
\end{figure}
\begin{table}
\centering
\begin{tabular}{|c|c|c|c|c|}
  \hline
  $Q$ & 0 &1 & 2 &3 \\ \hline
  $k_F$ & No & 2.33984327 & 4.67974416 & 7.01962931 \\ \hline
\end{tabular}
\caption{The Fermi momentum for different $Q$
with $q=p=2$.} \label{tablekfp2}
\end{table}
\begin{figure}[ht]
\centering
\includegraphics[scale=0.8]{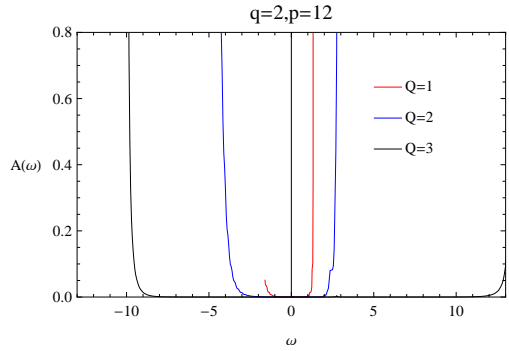}\hspace{1.5cm}
\caption{$A(\omega)$ for different values of $Q$
at zero temperature.} \label{A-Q}
\end{figure}

Now we will turn to investigate how the dilaton  field
influence the behavior of the spectral function.
Without loss of generality, we  set $q=p=2$ in
the discussion. When the dilaton field vanishes
$Q=0$,  the background (\ref{blackhole}) goes
back to the AdS Schwarzschild black hole. The
spectral function for $Q=0$ is shown in
Fig.\ref{figq1Q0p}. This background is not
charged, so that it does not have Fermi-like
peak.

In the dilaton gravity, the chemical potential is
finite as described in (\ref{tmu}). This gives
the possibility of fermionic excitation. The
numerical results on the spectral function for
nonzero values of $Q$ are shown in
Fig.\ref{densitydiffQ}. We observe the Fermi-like
peak and  present the Fermi momentum  in
Table.\ref{tablekfp2}. The Fermi momentum
increases but the peak becomes lower as the
increase of the dilaton field $Q$. For larger
$Q$, the flat band gets more dispersed in the low
momentum and the peak of the band appears at
higher momentum. This shows that the dilaton
field has the effect on the shift of the flat
band.

To see clearly the  effect of the dilaton field
on the gap, we plot the density of state by
integrating $A(\omega,k)$ over $k$ with strong dipole
coupling  $p=12$ in Fig.\ref{A-Q}. We observe
that with the increase of  $Q$, the gap becomes
wider. This tells us that in the dilaton gravity
the effect of the dipole coupling will be more
explicit.
\begin{figure}[ht]
\centering
\includegraphics[scale=0.6]{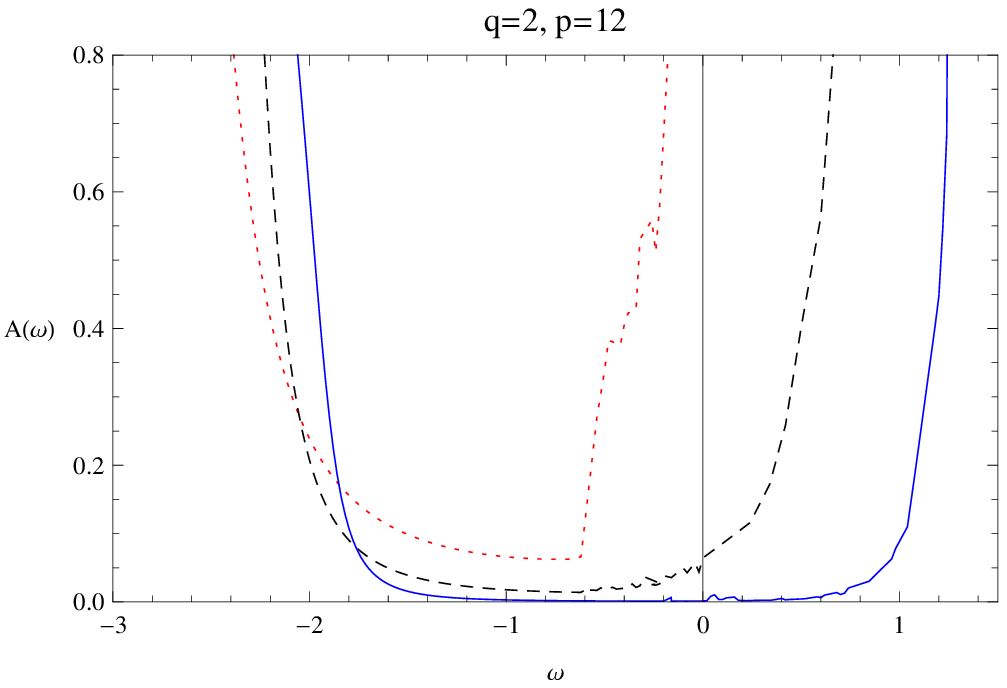}\hspace{1.5cm}
\includegraphics[scale=0.6]{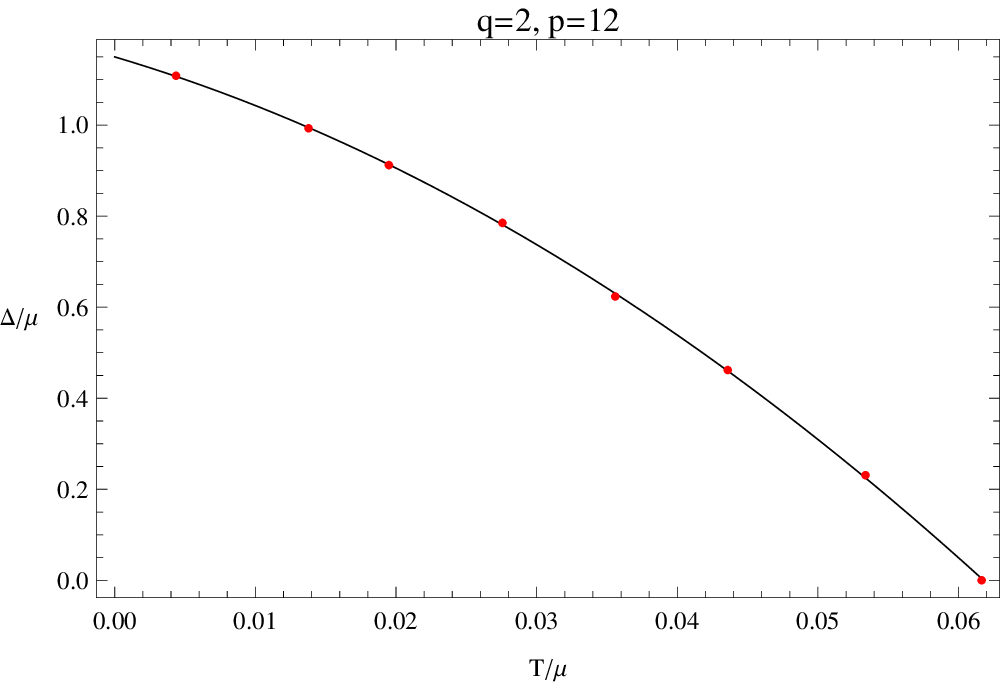}\hspace{1.0cm}
\caption{Left: $A(\omega)$ as a function of
$\omega$ when $Q=1$ at finite temperatures.
$T/\mu$  are $0.0435864$(solid),
$0.0616404$(dashed) and $0.0754938$(dotted).
Right: The width of the gap as a function of the
temperature.} \label{A-T1}
\end{figure}
\begin{figure}[ht]
\centering
\includegraphics[scale=0.6]{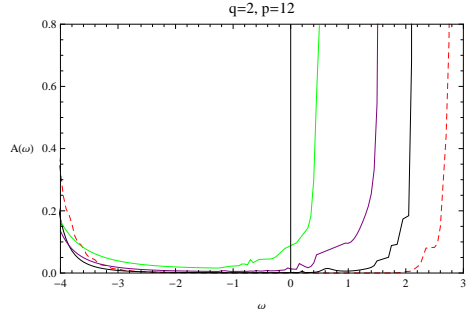}\hspace{1.5cm}
\caption{$A(\omega)$ as a function of $\omega$
when $Q=2$ at finite temperatures. $T/\mu$ are
$0$(red dashed), $0.0435864$(black),
$0.0533822$(purple) and $0.0616404$(green),
respectively.} \label{A-T2}
\end{figure}

\subsection{Finite temperature}
In \cite{R.G.Leigh2,Wen}, it was argued
that the gap brought by strong  dipole coupling
will vanish once the temperature increases to a
critical value for the relativistic fermionic
system. It is of interest to examine this
property for the non-relativistic situation
 by imposing Lorentz
breaking boundary conditions.

We calculate the spectral function at the finite
temperature. We set $q=2$ and $p=12$ in the
following discussion.  In the left panel of
Fig.\ref{A-T1}, we show the density of state for
different temperatures when the dilaton $Q=1$. We
see that the gap becomes narrower and closes up
when  $T/\mu$ increases to the value $0.061$.
Moreover, we show the width of the gap $\Delta$
as a function of $T$ in the right panel of
Fig.\ref{A-T1}. As the temperature increases to
$T_{\star}$ when the gap closes up, there is a
transition from the insulator to conducting
state. In our model, We find the ratio
$\Delta_{\star}/T_{\star}$ is around $18.65$,
where $\Delta_{\star}$ is the width of the gap at
zero temperature. This supports that the
generated gap is temperature-dependent.

We also calculate the spectral function for
different dilaton charge, for example taking
$Q=2$. The density of state for different
temperature is shown in Fig.\ref{A-T2}. The
property that gap becoming narrower for higher
temperature holds as well. For bigger dilaton
charge, we find that the value of
$\Delta_{\star}/T_{\star}$ is higher, where
$T_{\star}$ is the temperature  to close up the
gap. When $Q=2$,  the value of
$\Delta_{\star}/T_{\star}$  is around $27.82$.
This ratio  increases with the dilaton charge.
Note that in the charged AdS black hole
background, the ratio $\Delta_{\star}/T_{\star}$
was found around $10$ \cite{R.G.Leigh2}.  It is
worth pointing out that similar to the case of
zero temperature, the gap becomes wider as we
increase $Q$ for fixed finite temperature, this
is explicit by comparing Fig.\ref{A-T1} and
Fig.\ref{A-T2}.

\section{conclusion and discussion}
In this paper, we have studied extensively the properties of the holographic non-relativistic fermionic spectral function in the
presence of a bulk dipole coupling in the charged dilatonic black hole background.
We generalized previous related studies \cite{WJL2,Wen} by combining non-relativistic fermions,
dipole coupling and a dilaton field and studied how all these ingredients affect together the properties of this fermionic system.

Firstly, we have further confirmed that the emergence of flat band is robust in the case of non-relativistic fermionic fixed point,
independent of the black hole background and the coupling between the fermions and gauge field.
Secondly, we observed that the Fermi momentum increases as the dipole coupling becomes stronger for some specific charge $q$ and small $p$. By studying the dispersion relation for small $p$, we found that this fermionic system had the linear dispersion relation
for these specific charge $q$ and $p$. Then, by studying the large dipole coupling effects, we found that
for a fixed $q$, the Fermi sea disappeares and a gap opens up when the dipole coupling goes beyond a critical value,
which indicated this fermionic system possesses the characteristic of the Mott insulator. Our calculations also showed to us that bigger charge make the gap opens easier. Subsequently, we also investigated how the dilaton field influence the behavior of the spectral function. The results showed that the effect of the dipole coupling is more explicit in the dilaton gravity.
Finally, we found that this holographic non-relativistic fermionic system with dipole coupling
also exhibits a phase transition from insulator to a conducting state as the temperature is increased.
It implied that the gap emerged by the dipole coupling in this dilaton background possesses non-trivial temperature dynamics.

The properties for the nonrelativistic fermion
system disclosed here in the holographic study
with the modification of the boundary term are
interesting.  It is of interest to further
explore the holographic properties of the
nonrelativistic fermion system, for example the
backreaction of the bulk fermion on the background,
the nature of the Fermi surface and its relation
to the phase transition etc. Besides, it will be
interesting to examine whether the model
satisfies the Luttinger's
theorem\cite{Luttinger1,Luttinger2,Luttinger3}.
It is expected that the theoretical attempts in
this direction can help to unlock more puzzles in
the condense matter physics.

\begin{acknowledgments}
X. M. Kuang and B. Wang are supported partially by the NNSF of
China and the Shanghai Science and Technology
Commission under the grant 11DZ2260700. J. P. Wu is
partly supported by the National Research Foundation
of Korea(NRF) grant funded by the Korea government(MEST)
through the Center for Quantum Spacetime(CQUeST) of Sogang
University with grant number 2005-0049409 and also by the
NSFC under grant No.11275208.
\end{acknowledgments}


\begin{thebibliography}{99}

\bibitem{Hartnoll}
S.A. Hartnoll, Class. Quant. Grav. 26, 224002 (2009),[arXiv:0903.3246].

\bibitem{herzog}
 C.P. Herzog, J. Phys. A 42, 343001 (2009), [arXiv:0904.1975].

\bibitem{horowitz}
G.T. Horowitz, [arXiv: 1002.1722].

\bibitem{BHbackground1}
S. S. Gubser, F. D. Rocha, Phys. Rev. D
\textbf{81}, 046001 (2010), [arXiv:0911.2898];
\bibitem{BHbackground2}
K. Goldstein, S. Kachru, S. Prakash and S. P.
Trivedi, JHEP 1008 (2010) 078, [arXiv:0911.3586].
\bibitem{BHbackground3}
C. Charmousis, B. Gouteraux, B. S. Kim, E. Kiritsis and R. Meyer,
JHEP 1011 (2010) 151. [arXiv:1005.4690].

\bibitem{Huijse}
S. A. Hartnoll, L. Huijse, Class.Quant.Grav. 29 (2012) 194001, [arXiv:1111.2606].

\bibitem{D.Tong1}
J. N. Laia, D. Tong, JHEP 1111 (2011) 125, [arXiv:1108.1381].
\bibitem{D.Tong2}
J. N. Laia, D. Tong, JHEP 1111 (2011) 131 [arXiv:1108.2216]
\bibitem{WJL1}
W. J. Li, H. Zhang, JHEP1111(2011) 018, [arXiv:1110.4559].
\bibitem{WJL2}
W. J. Li, R. Meyer, H. Zhang, JHEP1201 (2012) 153, [arXiv:1111.3783].
\bibitem{JPWu3}
W. J. Li, J. P. Wu, Nuclear Physics B, 867 (2013), 810-826, [arXiv:1203.0674]

\bibitem{Lee}
S. S. Lee, Phys. Rev. D 79, 086006, (2009), [arXiv:0809.3402].

\bibitem{HongLiuUniversality}
N. Iqbal and H. Liu, Phys. Rev. D 79, 025023 (2009), [arXiv:0809.3808].

\bibitem{HongLiuADS2}
T. Faulkner, H. Liu, J. McGreevy and D. Vegh, Phys. Rev. D 83:125002,2011, [arXir:0907.2694].

\bibitem{HongLiuNonFermi}
H. Liu, J. McGreevy and D. Vegh, Phys. Rev. D 83, 065029 (2011), [arXiv:0903.2477].

\bibitem{HongLiuSpinor}
N. Iqbal and H. Liu, Fortsch. Phys. 57, 367 (2009), [arXiv:0903.2596].

\bibitem{Cubrovic}
M. Cubrovic, J. Zaanen and K. Schalm, Science 325 (2009) 439, [arXiv:0904.1993].

\bibitem{wujianpin}
J. P. Wu, JHEP 1107:106,2011, [arXiv:1103.3982].

\bibitem{Fang}
 L. Q. Fang, X. H. Ge, X. M. Kuang, Phys. Rev. D 86, 105037 (2012), [arXiv:1201.3832].

\bibitem{R.G.Leigh1}
M. Edalati, R. G. Leigh, P. W. Phillips, Phys. Rev. Lett. 106, 091602 (2011), [arXiv:1010.3238].
\bibitem{R.G.Leigh2}
M. Edalati, R. G. Leigh, K. W. Lo, P. W. Phillips, Phys. Rev. D 83, 046012 (2011), [arXiv:1012.3751].
\bibitem{JPWu2}
J. P. Wu, H. B. Zeng, JHEP 1204 (2012) 068, [arXiv:1201.2485]
\bibitem{Wen}
W. Y. Wen, S. Y. Wu, Phys. Lett. B712 (2012) 266-271, [arXiv:1202.6539].
\bibitem{Kuang}
X. M. Kuang, B. Wang, J. P. Wu, JHEP 07 (2012) 125, [arXiv:1205.6674].

\bibitem{Conventions1}
R. M. Wald, ``General Relativity", The University of Chicago Press.

\bibitem{Mitra}
S. S. Gubser and I. Mitra, JHEP 08 (2001) 018, [arXiv:hep-th/0011127].

\bibitem{Cvetic}
M. Cvetic, M.J. Duff, P. Hoxha et. al.,  Nucl. Phys. B558 (1999) 96 126, [arXiv:hep-th/9903214].

\bibitem{Photoemission}
T. Faulkner, G. T. Horowitz, J. McGreevy, M. M. Roberts, D. Vegh, JHEP 1003, 121 (2010), [arXiv:0911.3402].

\bibitem{McGreevy}
 D. Guarrera, J. McGreevy, [arXiv:1102.3908].

\bibitem{Muck}
W. M$\ddot{u}$ck, K. S. Viswanathan, Phys. Rev. D 58:106006,1998, [arXiv:hep-th/9805145].

\bibitem{J.Ren}
S. S. Gubser, Jie Ren, Phys. Rev. D 86:046004,2012, [arXiv:1204.6315].

\bibitem{Luttinger1}
N. Iqbal and H. Liu, Classical Quantum Gravity 29, 194004 (2012), [arXiv:1112.3671].

\bibitem{Luttinger2}
S. Sachdev, Phys. Rev. D 84, 066009 (2011), [arXiv:1107.5321 ].

\bibitem{Luttinger3}
A. Allais, J. McGreevy, and S. J. Suh, Phys. Rev. Lett. 108,231602 (2012), [arXiv:1202.5308].
\end{thebibliography}
\end{document}